\documentclass[a4paper,12pt,notoc]{JHEP3}
\usepackage[english]{babel}
\usepackage{cite}
\usepackage{graphicx,epsfig,bbm}

\newcommand{\be}{\begin{eqnarray}}
\newcommand{\en}{\end{eqnarray}}
\newcommand{\nn}{\nonumber}

\title{Phenomenology of symmetry breaking from extra dimensions}
\author{J. Alfaro, \\ Facultad de F\'{\i}sica, Pontificia Universidad Cat\'olica de Chile,
                      Casilla 306, Santiago 22, Chile}
\author{A. Broncano, \\ Max Planck Institute for Physics, F\"ohringer Ring 6, 80805 Munich,
                        Germany}
\author{M.B. Gavela, \\ Departamento de F\'{\i}sica Te\'{o}rica  and Instituto
                        de F\'{\i}sica Te\'{o}rica , Universidad Aut\'{o}noma de Madrid,
                        Cantoblanco, E-28049 Madrid, Spain}
\author{S. Rigolin, \\ Departamento de F\'{\i}sica Te\'{o}rica  and Instituto
                        de F\'{\i}sica Te\'{o}rica , Universidad Aut\'{o}noma de Madrid,
                        Cantoblanco, E-28049 Madrid, Spain}
\author{M. Salvatori \\ Departamento de F\'{\i}sica Te\'{o}rica  and Instituto
                        de F\'{\i}sica Te\'{o}rica , Universidad Aut\'{o}noma de Madrid,
                        Cantoblanco, E-28049 Madrid, Spain}
\preprint{FTUAM 06-6 \\ IFT-UAM/CSIC-06-24 \\ MPP-2006-49}
\abstract{
Motivated by the electroweak hierarchy problem, we consider  theories with two 
extra dimensions in which the four-dimensional scalar fields are components of 
gauge boson in full space. We explore the Nielsen-Olesen instability for $SU(N)$ 
on a torus, in the presence of a stationary background with constant field strenght.
A field theory approach is developed, computing explicitly the minimum of the 
complete effective potential, including tri-linear and quartic couplings and 
determining the symmetries of the stable vacua. We also develop appropriate 
gauge-fixing terms when both Kaluza-Klein and Landau levels are present and 
interacting, discussing the interplay between the possible six and four 
dimensional choices. The equivalence between coordinate dependent and constant 
Scherk-Schwarz boundary conditions -associated to continuous and/or discrete 
Wilson lines- is analyzed. }
\begin{document} 
%
\section{Introduction}

Data indicate that the mass of the Higgs boson is of the order of
the electroweak scale, $v\sim {\mathcal{O}}(100)$GeV. Such a mass is
unnaturally light if there is new physics beyond the Standard Model
(SM) and at a higher scale to which the Higgs boson is sensitive.
Generically, the Higgs mass is not protected by any symmetries and
thus gets corrections which are quadratically dependent on the new physics 
scale. The phenomenological success of the SM puts a lower bound on 
that hypothetical scale of about a few TeV~\cite{lep}, and it can even be as 
large as that at which quantum gravity effects appear, the Planck scale $M_{Pl}$.

Different scenarios have been devised to eliminate the quadratic
sensitivity of the Higgs mass to the cutoff scale, including: 
Higgs as a superpartner of a fermion and thus its mass is only
logarithmically ultraviolet (UV) divergent ({\it supersymmetry}), or
as a Goldstone boson of a spontaneously broken symmetry ({\it technicolor} 
\cite{technicolour} and {\it little Higgs} \cite{littlehiggs}), or as 
a component of a higher dimensional gauge multiplet ({\it gauge-Higgs 
unification} \cite{Manton,Hosotani, Hatanaka,Randjbar}). Independently of the 
precise nature assumed for the Higgs field, all these proposals require, 
in one way or another, the appearance of new physics at about the TeV 
scale. While the first two approaches are being intensely studied, in
practice they tend to be afflicted by rather severe fine-tuning
requirements when confronted with present data \cite{FineTuning}. In
this work, we concentrate on the last and less explored possibility
\cite{recent}.

We thus consider  theories formulated in more than four space-time
dimensions, with the extra dimensions compactified on tori of
generic length $L$, such that $v \ll 1/L \ll M_{Pl}$.  The idea is
that a single higher dimensional gauge field gives rise to the
four-dimensional ($4D$) fields: the gauge bosons, from the ordinary
space-time components, and the scalars, from the extra ones; the
Higgs field should then be identified among the scalars. The
essential point is that, although the $6D$ gauge symmetry is
broken by compactification, it remains locally unbroken.
Any local - sensitive to the UV physics - mass term for the scalars
is then forbidden and the Higgs mass would then have a non-local -
UV finite - origin.

Chiral fermions are an essential ingredient to achieve realistic
$4D$ effective models from higher-dimensional theories. This
requires the introduction of new ingredients in the above scenario.
Two main mechanisms have been explored for chirality:
\begin{itemize}
\item {\it Compactification on orbifold}
\cite{orbifold},  in which the extra dimensions are compactified on 
flat manifolds with singular points.
\item {\it Compactification with a background field}, either a scalar field
({\it domain wall} scenarios) \cite{Rubakov}, or gauge - and
eventually gravity - backgrounds with non trivial field strength
({\it flux compactification}) \cite{Randjbar}.
\end{itemize}

The idea of obtaining chiral fermions in
presence of abelian gauge and gravitational backgrounds was first  
proposed by Ranjbar-Daemi, Salam and Strathdee 
\cite{Randjbar}, on a $6D$ space-time with the two extra dimensions 
compactified on a sphere. This seminal idea was also retaken in string 
theory, more concretely in the heterotic string constructions \cite{Gross}.

The avenue explored in this work falls in this category: {\it flux compactification}, 
that is, compactification in  the presence of a gauge background with constant field 
strength. In this class of models, the mass splitting between the two chiralities 
is proportional to the field-strength of the stable background. That field strength 
vanishes on a two torus $\mathcal{T}^{\mathrm{2}}$ for simply connected groups such 
as $SU(N)$, precluding chirality in them. It may be non-zero instead for non-simply 
connected groups.  
 
A simple example would be to consider a $U(N)$ theory on $\mathcal{T}^{\mathrm{2}}$. 
As it is well known, the presence of a stable magnetic background associated with the 
abelian subgroup $U(1) \in U(N)$ induces chirality. Furthermore, it affects the 
non-abelian subgroup $SU(N) \in U(N)$, giving rise to a non-trivial \textit{t' Hooft 
non-abelian flux} \cite{tHooft79}. The latter induces rich symmetry breaking patterns. 
Notice that an analysis of $SU(N)$  is interesting in itself as regards the Higgs 
mechanism, as the Higgs field needs to have a non-abelian gauge parenthood in extra 
dimensions. 

Chirality from a gauge background can be seen as an hyperfine splitting induced by the 
field strength. A field theory treatment implies to solve the system in terms of fields 
which are charged or neutral with respect to the background, that is, in terms of Landau 
and Kaluza-Klein levels, respectively. It is interesting to develop the tools for such 
a field theory analysis, as they will be required to analyze the symmetry breaking patterns 
of general non-simply connected groups.
           
An historical field theory example of a theory involving both Kaluza-Klein and Landau 
levels is the analysis of the so-called {\it Nielsen-Olesen instability}\cite{Nielsen}. 
They studied a scenario within only the four usual flat dimensions, in order to 
justify confinement in QCD. A $SU(2)$ gauge theory in four dimensions was considered, 
with a background with constant field strength, that lived only on two of them and 
pointed to a fixed direction in the adjoint representation. They found that it resulted 
in an effective 2-dimensional $U(1) \in SU(2)$ invariant theory, including a scalar 
potential with charged (Landau like) and  neutral  (Kaluza-Klein like) fields. In the 
absence of such background, the lightest two charged ``scalars'' would be degenerate. 
In its presence hyperfine splitting follows automatically, though, with those two 
scalars acquiring squared-masses which are opposite in sign. One of the masses is 
tachyonic and thus may induce spontaneous symmetry breaking ``for free'': the $U(1)$ 
symmetry may be there but hidden. Such phenomenon is called in the literature 
\textit{Nielsen-Olesen instability}.
The meaning of the background and the subsequent instability, in the context of 
four infinite dimensions, is still a very controversial problem in the literature 
\cite{Leutwyler}.

In the present work, we solve the \textit{Nielsen-Olesen instability} for a $SU(N)$ gauge theory 
on $\mathcal{M}_4 \times \mathcal{T}^{\mathrm{2}}$. That is, we analyze the symmetry breaking 
induced by the presence of a background on the torus, which has constant field strength. The latter 
is assumed to point along a fixed direction of the adjoint representation and to be a function of 
the 't Hooft non-abelian flux. Notice, indeed, that although a constant field-strength is a solution 
of the equations of the motion, it is not necessarily a minimum of the action and may give rise to 
the presence of tachyonic degrees of freedom: the \textit{Nielsen-Olesen instability}.   
 
It is intriguing to consider whether the Nielsen-Olesen mechanism can be implemented for the purpose 
of electroweak symmetry breaking. Instead of enlarging the system so as to cancel {\it ab initio} 
any possible tachyonic term \cite{Cremades}, we explore here how a stable vacuum is reached from 
the initial configuration and we determine its remaining symmetries, for the simple toy model in 
consideration. Our target is to understand  from the field theory point of view the resulting 
four-dimensional scalar and vector sector and their symmetries. The field theory tools developed 
in this work will be useful and necessary in the future, when considering general non-simply connected 
gauge groups and/or higher dimensional (extra-dimensional) manifold.

Explicit field theory analysis of the minima of the effective four-dimensional 
Lagrangian in the presence of backgrounds have been attempted in the literature 
\cite{Nielsen, Antoniadis} for $SU(2)$, although in a rather incomplete way, due 
to the technical difficulties associated to handling simultaneously Kaluza-Klein 
and Landau levels in interaction. In contrast, we will take into account the 
complete effective $4D$ potential for the case of $SU(2)$, including all trilinear 
and quartic interaction terms. This will require to find a gauge-fixing Lagrangian 
appropriate when interacting towers of Kaluza-Klein and Landau levels are present, 
a tool not previously developed in the literature. As it will be shown, the 
six-dimensional $R_\xi$ gauge does not correspond to the four-dimensional one.
Furthermore, it will be technically necessary to solve integrals involving two, 
three and four Kaluza-Klein and Landau levels: this will be done analitically for 
all modes. In the present case, they will allow us to compute 
the four-dimensional potential, find its minima and determine then the spectra 
and their symmetries.
These technical results could be useful in more general scenarios 
than those considered here. For example, it has been suggested that unstable flux configurations 
can be associated with unstable intersecting branes configurations \cite{Taylor:1997dy}. 
In this context, our field theory approach can be seen as a classical approximation 
of a D-brane decay via open-string tachyon condensation \cite{Sen:2004nf}.


Were $SU(N)$ the interesting gauge group, the field theory treatment described above 
would have been unnecessary, as pure theoretical arguments allow to argue the 
symmetries of the stable vacua. On $\mathcal{T}^{\mathrm{2}}$, a background with 
constant field strength requires coordinate-dependent boundary conditions for the 
fields. For the particular case of the gauge group $SU(N)$, they are gauge equivalent 
to constant boundary conditions \cite{Ambjorn,Salvatori}. The symmetries of the 
four-dimensional spectra can thus be inferred. The vacuum symmetries depend 
essentially on whether trivial or non-trivial `t Hooft fluxes are present, which 
translates then on whether the constant boundary conditions correspond to continuos 
or continuos and discrete Wilson lines, respectively. While much literature is dedicated 
to the case of continuos Wilson lines, one of the novel ingredients of this paper is the 
phenomenological analysis of the pattern of gauge symmetry breaking and the spectrum of 
four-dimensional gauge and scalar excitations, for the general case of $SU(N)$ with 
non-trivial 't Hooft flux. The results will be shown to be consistent with those 
obtained from the field theory analysis of the effective Lagrangian, for the case 
of $SU(2)$, further supporting the consistency of the field theory approach developed 
in this work.
 
In Section $2$, general theoretical arguments prove the existence of
absolute minima, for $SU(N)$. Boundary conditions depending on the extra 
coordinates are shown to be equivalent to constant ones and the expected 
symmetry breaking patterns for the stable vacua are determined. In Section 
$3$ the problem is reformulated in terms of the $6D$ $SU(N)$ Lagrangian. Next 
we obtain the complete effective four-dimensional Lagrangian out of the 
explicit integration of the $6D$ Lagrangian over the torus surface, for 
the $SU(2)$ case; appropriate gauge-fixing conditions are proposed and 
developed in detail as well. In Section $4$ the stable minima of the 
complete four-dimensional potential and the resulting physical spectra is 
identified, for the $SU(2)$ case. The last step of this procedure is
done numerically and the results are then compared with the symmetry
breaking patterns expected from the general theoretical analysis 
developed in Section $2$. In Section $5$ we conclude. The Appendices 
contain supplementary arguments and develop further technical tools.

\section{Vacuum energy}
\label{vacuum}

Consider a $6D$ $SU(N)$ gauge theory, with generators $\lambda^a$ defined by
Tr$[\lambda^a \lambda^b]=\delta^{ab}/2$ and $[\lambda^a,\lambda^b]= i f^{abc} 
\lambda^c$. The Yang Mills Lagrangian reads
\be
 \mathcal{L}_6 = - \frac{1}{2} \, {\rm Tr}[\mathbf{ F}_{M N} \mathbf{F}^{MN} ] =
 - \frac{1}{4} \, \mathbf{F}_{MN}^{a} \mathbf{F}^{MN}_{a} \;,
\label{L6compact}
\en
where
\be
\mathbf{F}_{MN}^{a} &=& \partial_M \mathbf{A}_{N}^a - \partial_N \mathbf{A}_M^a
 + g  f^{abc} \mathbf{A}_M^{b} \mathbf{A}_N^c \;,
\en
and $\mathbf{A}_M^a$  are the gauge fields in the adjoint representation of the group. 
Throughout the paper,  Greek (Latin) indices will denote the ordinary (extra) coordinates. 
The two extra dimensions are compactified on an orthogonal torus ${\mathcal T}^2$, with
compactification lengths $l_1$, $l_2$, and area $\mathcal{A}=l_1 l_2$. In what follows, 
we will denote by $x$ the four Minkowski coordinates and by $y$ the two extra coordinates.

We assume a constant field strength pointing to an arbitrary direction in gauge space. 
We also assume $4D$ Poincar\'e invariance. In accordance with it, the background 
can only be of the form $B_M = \left(0,B_i^a(y)\right)$. The gauge fields can then be
parametrized in terms of that classical background, $B_M^a$, and the 
fluctuations $A_M^a$,
\be
\mathbf{A}_M^a (x, y) =
B_M^a(y)+ A_M^a(x,y)\;,
\label{shiftbac}
\en
allowing to decompose the total field strength as
\be
\mathbf{F}_{MN}^a(x,y)=
G_{MN}^a\,+\,F_{MN}^a(x,y)\,,
\label{Fexpansion}
\en
with $G_{MN}$ given by
\be
G^a_{\mu\nu}= 0\,,\qquad
G^a_{\mu i}= 0\,,\qquad
G_{ij}^{a} = \partial_i B_{j}^a - \partial_i B_j^a + g  f^{abc} B_i^{b} B_j^c \,.
\label{tata}
\en
In what follows, $B_i(y)$ and $G_{ij}$ will be denoted {\it imposed} background and field 
strength, respectively, which do not necessarily coincide with those of a true -stable- 
vacuum configuration. The latter will be instead dubbed {\it total}.

To live on a torus implies to specify boundary conditions, which describe how fields 
transform under translations by $l_1$ and $l_2$. Let $T_i$ be the embedding of such 
translations in gauge space. Upon their action, gauge fields in the adjoint representation 
can vary at most by a gauge transformation,
\be
\label{perio-su2-gauge}
\mathbf{A}_M (x,y+l_i) = T_i(y) \,\mathbf{A}_M (x,y) \,T_i^{\dagger}(y) + 
                        \frac{i}{g}\, T_i (y)\,\partial_M T_i^{\dagger} (y)\,.
\label{T}
\en
Translations $T_i$ must, in general, commute up to an element of the center of the group,
\be
T_2^{-1}(y_1,y_2)\,T_1^{-1}(y_1,y_2+l_2)\,T_2(y_1+l_1,y_2)\,T_1(y_1,y_2)=
e^{2\pi i (k+\frac{m}{N})}\,,
\label{conmut}
\en
where $k$ and $m$ are integers, with $m$ being the {\it 't Hooft non-abelian flux} \cite{tHooft79}, 
a gauge invariant quantity constrained to take values between $0$ and $(N-1)$.

Given a set of $T_i$, the possible backgrounds $B_i$ are constrained by Eq.~(\ref{T}), implying
\be
\label{trans1}
{A}_M (x,y+l_i) &=& T_i(y) \,{A}_M (x,y) \,T_i^{\dagger}(y)\,,\\
{F}_{MN} (x,y+l_i) &=& T_i(y) \,{F}_{MN} (x,y) \,T_i^{\dagger}(y)\,, \\ 
{B}_j (y+l_i)   &=& T_i(y) \,{B}_j (y) \,T_i^{\dagger}(y) + 
                    \frac{i}{g}\, T_i (y)\,\partial_j T_i^{\dagger} (y)\, ,\\
{G}_{MN} &=& T_i(y) \,{G}_{MN} \,T_i^{\dagger}(y)\,.
\en

\subsubsection*{Instability}
For a $SU(N)$ theory on a two-dimensional torus, an expansion around a constant field 
strength corresponds to a background configuration that satisfies the equations of 
motion, but it is not stable. A simple argument goes as follows. 
Given a constant $G_{12}$, the only mass term present in the $6D$ Lagrangian for the 
$6D$ field excitations is
\be
  - g  f^{abc} A_1^{b} A_2^c\,  G^{12}_a\,.
\label{unst}
\en
Because the background field strength $G_{12}$ is a non-zero Lorentz constant, the 
anticommutativity of $f^{abc}$ implies then the presence in the Lagrangian of a 
field with negative mass, as can be seen rewriting Eq.~(\ref{unst}) in the diagonal 
basis\footnote{Other possible mass terms, resulting after fixing the gauge for the 
excitation fields, only produce symmetric terms, which cannot cancel the antisymmetric 
contributions in Eq.~(\ref{unst}).}. In other words, the mass matrix defined by 
Eq.~(\ref{unst}) is a traceless quantity and, for $G_{12} \neq 0$, it necessarily has 
at least one positive and one negative mass eigenvalue\footnote{This is unlike the $U(N)$ 
case, for instance, where the $U(1)$ part is not subject to such a constraint.}. 

The instability argument for a background with constant field strength can be also discussed  from a $4D$
point of view. The $4D$ Lagrangian is
  \be
  \mathcal{L}_4 & = & \int_{\mathcal{T}^2} \hspace{-0.25cm } d^2 y \, \mathcal{L}_6
                  = -\frac{1}{2} \int_{\mathcal{T}^2} \hspace{-0.25cm } d^2 y \,
                    {\rm Tr}\, [\mathbf{F}_{MN} \mathbf{F}^{MN}] = \nn \\
                & = & -\frac{1}{2}\,\int_{\mathcal{T}^2} \hspace{-0.25cm } d^2 y \,
                     {\rm Tr}\, [\mathbf{F}_{\mu \nu}\mathbf{F}^{\mu \nu} +
                     2 \mathbf{F}_{\mu i} \mathbf{F}^{\mu i} + \mathbf{F}_{ij} \mathbf{F}^{ij}]\, .
   \label{lagexp}
   \en
Our aim is to identify the possible degenerate vacuum solutions consistent with
$\mathbf{F}_{\mu \nu} \mathbf{F}^{\mu \nu}=0$ and compatible with the boundary conditions.
$4D$ Lorentz and $4D$ translation invariance on a flat  ${\mathcal M}_4\times \mathcal{T}^2$ 
manifold also require that, at the minimum, $\mathbf{F}^{\mu i} = 0$. The third term in
Eq.~(\ref{lagexp}) is positive semi-definite,
\be
\int_{\mathcal{T}^2} \hspace{-0.25cm } d^2 y \, {\rm Tr}\, [\mathbf{F}_{ij}^2]\,\geq\,0 \,.
\label{min2}
\en
For a $SU(N)$ gauge theory on a $2D$ torus, the energy is not bounded from below by 
any topological quantity\footnote{Notice the difference between $SU(N)$ and $U(N)$ on 
$T_2$. In $U(N)$, $\int_{\mathcal{T}^2} {\rm Tr}\, [ \mathbf{F}_{ij}^2] \ge 
(1/4)\int_{\mathcal{T}^2}| {\rm Tr}\, (\epsilon_{\mu \nu} F^{\mu \nu})|^2$, 
which may be non-zero.}. Consequently, the absolute minimum should correspond to 
the lower limit of the inequality Eq.~(\ref{min2}), implying
\be
\mathbf{F}_{ij}^a\big|_{min}\equiv {\widetilde G}_{ij}^a\,=\,0,\,\,\,\,\,\forall \,i,\,j,\,a\hspace{1cm}\Rightarrow
\hspace{1cm} {\widetilde G}_{ij}^a= G_{ij}^a\,+\,F_{ij}^a\big|_{min}\,=0\,,
\label{absin}
\en
where Eq.~(\ref{Fexpansion}) has been used. In the above and from now on we denote with
$\sim$ the quantities pertaining to the {\it total} stable vacua, which has vanishing field 
strength, ${\widetilde G}_{ij}^a=0$.

In other words, the original {\it imposed} configuration, with constant  background field strength, 
$G_{ij}^a$, is not stable. In order to satisfy Eq.~(\ref{absin}) the scalars contained 
in the $4D$ potential,
\be
V = \frac{1}{2}\int_{\mathcal{T}^2} \hspace{-0.25cm } d^2 y \, {\rm Tr} [F_{ij}^2 + 2 \, G_{ij}\,F_{ij}]\,,
\label{vmin4D}
\en
will have to develop  vacuum expectation values, allowing the system to evolve towards 
a stable vacuum.  That is, it is to be expected 
that the system will respond to the {\it imposed} background through a pattern alike 
to that of $4D$ spontaneous symmetry breaking.

Furthermore, as the total vacuum energy will correspond to
\be
E_{tot} = \frac{1}{2} \int \hspace{-0.1cm} d^4 x \, \int_{\mathcal{T}^2} \hspace{-0.25cm } d^2 y\,\,
          {\rm Tr} [\mathbf{F}_{ij}^2\big|_{min}] = 0 \,,
\label{truevacuum}
\en
the absolute minima will have to be reached from the initial {\it imposed} background through
a pattern of scalar vacuum expectation values which, at the classical level, {\bf do not} 
contribute to the cosmological constant, which thus remains being zero.

\subsubsection*{The true vacuum}
The true vacuum should correspond to a configuration of zero energy, $\widetilde G_{MN}=0$, as 
explained above.
Let $\widetilde B_i (y)$ be such a stable background configuration, whose precise form remains
to be found. ${\widetilde B_i(y)}$ can be interpreted as the sum of the {\it imposed} background
$B_i(y)$ plus that resulting from the system response. A $SU(N)$ gauge configuration of zero
energy is a pure gauge and may be expressed by
\be
\widetilde B_i (y) = \frac{i}{g} U(y) \partial_i U^\dagger(y)\,,
\label{min}
\en
where $U$ is a $SU(N)$ gauge transformation. 
The problem of finding the non-trivial vacuum of the theory reduces, then, to build a $SU(N)$
gauge transformation $U(y)$ compatible with the boundary conditions. Substituting Eq.~(\ref{min})
into Eq.~(\ref{perio-su2-gauge}), it follows that $U$ must satisfy
\be
U(y+l_i) = \,T_i(y) \,U(y) \,V_i^\dagger \,,
\label{U_cons}
\en
where $V_i$ are arbitrary constant elements of $SU(N)$, only subject to the constraint
\be
 V_1^{-1} \,V_2^{-1} \,V_1 \,V_2\,=\,e^{2 \pi i (k+\frac{m}{N})}\,.
\label{V_cons}
\en
For $ SU(N)$ on a $2D$ torus, it is always possible \cite{Ambjorn,Salvatori} to solve recursively the
boundary conditions (\ref{U_cons}) and consequently such an $U$ exists.

Under a gauge transformation $S \in SU(N)$, the embeddings of translations transform as
\be
T_i' (y) = S(y+l_i) \, T_i(y) S^\dagger(y) \,.
\en
In order to catalogue the possible degenerate vacua, it is useful to work in a gauge that 
we will denote as $6D$-{\em background symmetric gauge}: that in which the {\it total} 
vacuum gauge configuration is trivial, ${\widetilde {B}}_M^{sym}=0$. Upon the gauge 
transformation  $S=U^\dagger$, with $U$  defined in Eq.~(\ref{U_cons}), it results
\be
T_i^{sym} = U^\dagger(y+l_i) T_i(y) U(y) = V_i \,,\hspace{1.0cm}
{\widetilde {B}}_M^{sym}=0\,.
\en
In this gauge the background is then zero and the constant matrices $V_i$ coincide 
with the boundary conditions. To classify the classical degenerate minima is then 
tantamount to classify the possible constant matrices $V_i$. The symmetries of the 
vacuum correspond to those generators commuting with all $V_i$. 
$V_i$ can be parametrized as 
\be
V_i \equiv e^{2 \pi i \alpha_i^a \lambda^a} \,,
\label{V_para}
\en
with  $\alpha_i^a$ being arbitrary constants only subject to the consistency condition 
(\ref{V_cons}). Two main cases can occur depending on whether the value of $m$ in 
Eq.~(\ref{conmut}) is equal to zero or not. Notice that:
\begin{itemize}
\item For $m=0$, as the embeddings of translations $V_i$ commute, it is possible to perform a non-periodic gauge transformation leading to   gauge  fields which transform ``periodically", while the boundary conditions are reabsorbed in the vacuum expectation values of scalar fields (Hosotani mechanism).
\item For $m\neq0$, on the contrary, as the $V_i$ do not commute,  such a transformation to periodic boundary conditions is not achievable.
\end{itemize}

\subsection{Trivial 't Hooft flux: $m=0$}

The name reminds that, in this case, the embedding of translations in gauge space 
commute and all classical vacuum solutions are degenerate in energy with the trivial 
vacuum, which is $SU(N)$ symmetric.

For $m=0$, the $V_i$ constant matrices commute, constraining the possible $\lambda^a$ 
in Eq.~(\ref{V_para}) to belong to the $(N-1)$ generators of the Cartan subalgebra. 
The vacua are thus characterized by $2(N-1)$ real continuous parameters $\alpha^a_i$,
$0 \leq \alpha^a_i < 1$. These $\alpha_a^i$ are non-integrable phases, which only arise 
in a topologically non-trivial space and cannot be gauged-away. Their values must be 
dynamically determined at the quantum level: only at this level the degeneracy among 
the infinity of classical vacua is removed \cite{Hosotani}.

The solution with $\alpha_i^a=0$ is the trivial, $SU(N)$ symmetric, one. For non-zero 
$\alpha_i^a$ values, the residual gauge symmetries are those associated with the 
generators that commute with $V_i$. As $V_1$ and $V_2$ commute, the rank of $SU(N)$ 
cannot be lowered \cite{Hebecker} and thus the maximal symmetry breaking pattern 
that can be achieved is
\be
SU(N) \longrightarrow U(1)^{N-1}.
\en
The spectrum of the $4D$ fields corresponding to the Cartan subalgebra
is that of an ordinary Kaluza-Klein (KK) tower,
\be
M^2_{n_1,n_2} = 4 \pi^2 \left[ \frac{n_1^2}{l_1^2} +  \frac{n_2^2}{l_2^2} \right]\,
,\,\,\,\,\,\,\, n_1,n_2 \in  {\mathbb Z}\,,
\label{KK}
\en
whereas for the rest of the fields, that is, fields corresponding to generators that 
do not commute with all $V_i$, the spectrum is expected to be of the form
\be
M^2_{n_1,n_2} = 4 \pi^2 \left[\frac{ (n_1 + \sum_{a=1}^{N-1} \,q^a \alpha_1^a/2\,)^2}{l_1^2} +
                              \frac{ (n_2 + \sum_{a=1}^{N-1} \,q^a \alpha_2^a/2\,)^2}{l_2^2} \right] \,,
\label{spectrum0}
\en
where $q^a$ are the field charges, expressed in units of the charge of the fundamental 
representation. These type of spectra are characteristic of Scherk-Schwarz symmetry
breaking scenarios \cite{SS,Luscher,Hosotani,Ferrara:1988jx}.

In the simplest case of $SU(2)$, that will be of interest for us in the following sections,
the two $V_i$ matrices may be chosen\footnote{The direction $a=3$ is only a possible choice; 
obviously the choice of gauge direction in the parametrization is arbitrary. It bears no 
relationship with the gauge direction chosen for the {\it imposed} background.} to be for 
instance $V_1=e^{\pi i\alpha_1\sigma_3} $ and $V_2=e^{\pi i\alpha_2\sigma_3}$.

The mass spectrum for the fields $A_M^3$ coincides with the KK spectrum (\ref{KK}), whereas 
for fields which do not belong to the Cartan subalgebra is is given by
\be
M^2_{n_1,n_2} = 4 \pi^2 \left[ \frac {(n_1 \pm \alpha_1 \,)^2}{l_1^2} +
                               \frac {(n_2 \pm \alpha_2 \,)^2}{l_2^2} \right] \,,
\label{spectrum0_2}
\en
as $q^a=2$ for fields in the adjoint representation. There are no massless modes in this sector,
for non-zero $\alpha_i$. The expected symmetry breaking pattern is thus
\be
SU(2) \longrightarrow U(1) \,.
\en

\subsection{Non-trivial 't Hooft flux: $m\neq 0$}

In this case, all solutions exhibit symmetry breaking, even at the classical level. 
The embeddings of translations in gauge space do not commute, Eq.~(\ref{conmut}), 
and the same holds then for the constant matrices $V_i$ \cite{tHooft79,Witten,callan}. 
In consequence, the symmetry breaking pattern lowers the rank of the group 
\cite{Guralnik,vanBaal:1985na,Salvatori} 

For $m\neq0$, Eq.(\ref{V_cons}) reduces to the so-called two-dimensional twist algebra \cite{tHooft79}. 
The first solutions were found in Refs.\cite{Ambjorn}. The problem for the four-dimensional case 
was addressed and solved in Refs.\cite{tHooft81,Gonzalez-Arroyo:1982ub,vanBaal:1982ag}. The most 
general solution up to four dimensions was obtained in Refs.\cite{vanBaal:1983eq,Brihaye:1983yd} 
and the $d$-dimensional case was studied in Refs. \cite{Gonzalez-Arroyo:1997uj,vanBaal:1985na}.

The irreducible representations of the two-dimensional algebra in Eq.(\ref{V_cons}) are given 
by \cite{vanBaal:1985na}
\be
\left\{
\begin{array}{ccc}
V_1 & = & \omega_1\,P^{s_1}\,Q^{t_1} \\
V_2 & = & \omega_2\,P^{s_2} \,Q^{t_2}  
\end{array} \right. \,,
\label{VPQ}
\en
where 
$P\equiv e^{i \pi (N-1)/N}\,diag( 1, e^{2 \pi i \frac{1}{N}}, ... , e^{2 \pi i \frac{N-1}{N}})$, 
$Q_{ij}\equiv e^{i \pi (N-1)/N}\,\delta_{ij-1}\,$, satisfying $P^N=Q^N=e^{i\pi(N-1)}$ and 
$ P\,Q\,=\,e^{2 \pi i/N}\,Q\,P$. The parameters $s_i,\, t_i$ are integers that assume values between 
$0$ and $N-1$ (modulo $N$) and that have to satisfy the consistency condition 
\be
s_1 \,t_2 \,\,-\,\,s_2\,t_1\,\,=\,\,m \,.
\label{cons_alp_beta_m}
\en
Let us define the quantity\footnote{g.c.d.= great common divisor} $\mathcal{K}=$ g.c.d.$(m,N)$. 
The constant matrices $\omega_1,\,\omega_2$ are elements of the subgroup $SU(\mathcal{K}) 
\,\subset \,SU(N)$ which satisfy the constraint: $\left[ \omega_1,
\omega_2\,\right] \,= \,0$. Since the $\omega_i$ commute, they can be parametrized in terms 
of generators, $H_j$, belonging to the Cartan subalgebra of $SU(\mathcal{K})$:
\be
\omega_i = e^{2 \pi i \sum_{j=1}^{\mathcal{K}-1} \alpha_i^j H_j} \,,
\en
where $\alpha_i^j$ are $2 (\mathcal{K}-1)$ real continuous parameters assuming values in the 
interval $0 \leq \alpha_i^j < 1$. 

The solutions of the algebra in Eq.(\ref{V_cons}), and consequently the classical vacua, are 
characterized by four integers, $s_i$ and $t_i$, and by $2 (\mathcal{K}-1)$ continuous parameters 
$\alpha_i^j$. However, all possible sets of $s_i, t_i$ can be seen as different parametrizations 
of the same vacuum \cite{vanBaal:1985na}: the classical vacuum is completely described only by 
$\mathcal{K}$ and the non-integrable phases $\alpha_i^j$. 
If $\mathcal{K}>1$, there is a degeneracy among an infinity of classical vacua and the true vacuum 
can be dynamically determined only at the quantum level \cite{Hernandez} as for the $m=0$ case. 
Only at this level it is possible to fix the values of $\alpha_i^j$ and to remove such degeneracy.

The pattern as well as the nature of symmetry breaking induced by the constant matrices $V_i$ in 
Eq.(\ref{VPQ}) is discussed in detail in Refs.\cite{Salvatori, Hernandez}. Here we only summarize 
the main results that are going to be used in the following sections. The symmetry breaking pattern 
turns out to be: 
\be
SU(N) \rightarrow SU(\mathcal{K}) \rightarrow \mathcal{H}_{(\mathcal{K}-1)} \,,
\label{sym_break}
\en
where $\mathcal{H}_{(\mathcal{K}-1)}$ denotes a subgroup of $SU(\mathcal{K})$ 
with the same rank. 

The first symmetry breaking step in Eq.(\ref{sym_break}) is due to the matrices $P$ and $Q$: since 
they don't commute, this symmetry breaking is rank-lowering and can be seen as explicit. In particular, 
if $\mathcal{K} = 1$, $SU(N)$ is completely broken. For $\mathcal{K}>1$, the presence of non-trivial 
$\omega_i$ allows a supplementary symmetry breaking, which is rank preserving and can be seen as spontaneous. 

It can be shown that the mass spectrum is arranged along towers of fields \cite{Salvatori} 
whose masses can be expressed as:
\be
({M^{a}_{n_1,n_2}})^2 = 4 \pi^2 \left[ \left(n_1 +  \beta^a_1 \right)^2\frac{1}{l_1^2} \,+ \,
                                       \left(n_2 +  \beta^a_2 \right)^2\frac{1}{l_2^2} 
                                \right] \,,
\label{spectrum1}
\en
where $\beta^a_i$ synthetically accounts for all possible contributions.
The components of $\beta^a_i$ responsible for the $SU(N) \rightarrow SU(\mathcal{K})$ 
symmetry breaking are quantized (i.e. $\beta^a_i=0,\mathcal{K}/N, 2 \mathcal{K}/N$,...,$1-\mathcal{K}/N$) as a consequence of the commutation 
rule between $P$ and $Q$. On the other side, the components of $\beta^a_i$ that induce the 
$SU(\mathcal{K}) \rightarrow \mathcal{H}_{(\mathcal{K}-1)}$ symmetry breaking 
are continuous parameters depending on the non-integrable phases $\alpha_i^j$. In summary, these type of 
spectra are characteristic of constant Scherk-Schwarz boundary condition scenarios, although with the 
contemporaneous presence of quantized and continuous parameters. 

As an illustration, let us particularize again to the $SU(2)$ case. The only possible 
non-zero value of $m$ is then $m=1$, for which a possible choice for the $P$ and $Q$ 
matrices is $P = i\sigma_3$ and $Q = i\sigma_1$, with $V_i$ given by
\be
\begin{array}{cc}
\left\{\begin{array}{c}
V_1 =i \sigma_3 \\
V_2 = i\sigma_1
\end{array} \right. & \qquad {\rm or} \qquad
\left\{\begin{array}{c}
V_1 = i\sigma_1 \\
V_2 = i\sigma_3
\end{array} \right.
\end{array}\,.
\label{su2_v}
\en
As $\mathcal{K}=1$, Eq.(\ref{sym_break}) entails that the expected symmetry breaking pattern is 
\be
SU(2) \longrightarrow \varnothing \,, \nn
\label{allbroken}
\en
even at the classical level. Three towers of fields result, with masses given by
\be
M_{n_1,n_2}^2=
\left\{
\begin{array}{l}
\displaystyle
4 \pi^2 \left[ \frac{\left(n_1 + 1/2\right)^2}{l_1^2} +
  \frac{n^2_2}{l_2^2} \right]  \\
\\  
\displaystyle
 4 \pi^2 \left[ \frac{\left(n_1 + 1/2\right)^2}{l_1^2} +
  \frac{\left(n_2 + 1/2\right)^2}{l_2^2} \right] \\
  \\
\displaystyle
 4 \pi^2 \left[ \frac{n^2_1}{l^2_1} +
  \frac{\left(n_2 + 1/2\right)^2}{l_2^2}  \right] \, .
 \end{array}\,
\right.
\label{spectrum1_2}
\en
These expressions allow no zero modes and thus the $SU(2)$ gauge symmetry is indeed 
completely broken\footnote{With the particular choice in Eq.~(\ref{su2_v}) the three 
towers in Eq.~(\ref{spectrum1_2}) would correspond to the gauge directions $a=1,2,3$, 
respectively.} \cite{Daniel,Gava}.

To conclude this Section, we have seen that for $SU(N)$ on a $2D$ torus, the $y$-dependent 
boundary conditions are equivalent to constant Scherk-Schwarz boundary conditions, $V_i$. 
For the case of trivial-'t Hooft flux, $m=0$, the treatment shows them to be equivalent to 
boundary conditions associated to continuous Wilson lines. For the non trivial 't Hooft case, 
$m \ne 0$, they are equivalent to boundary conditions associated contemporaneously to discrete 
and continuous Wilson lines. If $\mathcal{K}=1$ only discrete Wilson lines are present.

\section{The effective Lagrangian theory}
\label{6DL}

In the rest of the paper, we will analyze the pattern of symmetry breaking within a completely
different approach: the identification of the minimum of the effective $4D$ potential, after
integrating the initial $6D$ Lagrangian -with a constant background field strength- over the extra 
dimensions. To find and verify explicitly the form of the true vacuum, solving the Nielsen-Olesen instability on the torus, we will obtain the 
$4D$ scalar potential and minimize it. 
After some general considerations for $SU(N)$, we will treat in full detail the 
$SU(2)$ case and compare the resulting spectra with those predicted in the previous Section. 

\subsection{The $6$-dimensional $SU(N)$ Lagrangian}

The Yang-Mills Lagrangian Eq.~(\ref{L6compact}) can be rewritten in terms of the {\it imposed}
background and its fluctuations as
\be
\mathcal{L}_{YM}
&=& - \frac{1}{4} ( G_{MN}^a + F_{MN}^a)^2  =
\mathcal{L}^{(0)}_{A}+ \mathcal{L}^{(1)}_{A}+ \mathcal{L}^{(2)}_{A}+ \mathcal{L}^{(3)}_{A} + \mathcal{L}^{(4)}_{A}\;,
\en
where the Lagrangian terms corresponding to $i=0,1,2,3, 4$ fluctuation fields are, explicitly,
\be
\label{L0}
\mathcal{L}^{(0)}_{A} &=& -\frac{1}{4} \, G_{MN}^{a} G^{MN}_{a} \\
\label{L1}
\mathcal{L}^{(1)}_{A} &=& -\frac{1}{2} \, G_{MN}^a (D^M A^{N\,a} - D^N A^{M\,a})] \\
\label{L2}
\mathcal{L}^{(2)}_{A} &=&
-\frac{1}{2} [ D_M A_N^{a} \,D^M A^{N\,a} -D_M A_N^{a} \, D^N A^{M\,a} +  g f^{abc} G_{MN}^a A_b^M A_c^N ] \\
\label{L3}
\mathcal{L}^{(3)}_{A} &=&  -\frac{1}{2}\,g f^{abc}  (D^M A^{N\,a} - D^N A^{M\,a}) A^b_M A^c_N \\
\label{L4}
\mathcal{L}^{(4)}_{A} &=&   -\frac{1}{4} \,g^2 f^{abc} f^{amn} A_M^b A_N^c A^M_m A^N_n \;.
\en
The form of $G_{MN}$ was given in Eq.(\ref{tata}), while
\be
F_{MN}^a = D_M A_N^a - D_N A_M^a + g f^{abc} A_M^b A_N^c,
 \label{F}
 \en
 with $D_M$ being the {\it imposed}-background covariant derivative,
\be
D_M A_N^a & \equiv & \partial_M A_{N}^a  - g  f^{abc} A_N^{b} B_M^c \;, 
          \label{derivataback}
\en
satisfying
\be
\left[ D_M, D_N \right] & = & - i\, g\, G_{MN} \,. 
          \label{conm12}
\en
Notice that classically $\mathcal{L}^{(1)}_{A}=0$, as the {\it imposed} background satisfies 
the stationarity condition given by the equations of motion, ${D^a}_M G^{MN}=0$, although 
we will see below this it is not a stable vacuum configuration.

A possible choice for the {\it imposed} background, compatible with constant $G_{MN}$, is
\be
B_i(y) = - \epsilon_{ij} \,\frac{2\pi}{g} \left(k + \frac{m}{N}\right)
           \,\frac{y_j}{\mathcal{A}}\,\,\hat\lambda\,,
\label{back-suN}
\en
where $\hat\lambda$ denotes an arbitrary direction in gauge space, leading to
\be
G_{12}=\frac{4 \pi (k+\frac{m}{N})}{g\,\mathcal{A}} \, \hat\lambda\,\equiv \,\frac{2}{g}\, 
       \mathcal{H} \,\hat\lambda\, .
\label{G12}
\en
The quantity $\mathcal{H}$ so defined can be interpreted as a quantized abelian magnetic flux over the 
torus surface (up to some factors):
\be
\frac{1}{\mathcal{A}}\,\,\int_{\mathcal{T}^2} \hspace{-0.25cm } d^2 y \,
         \left( \partial_1 B_2 - \partial_2 B_1  \right)=
\frac{1}{\mathcal{A}}\,\,\int_{\mathcal{T}^2} \hspace{-0.25cm } d^2 y \,
         G_{12} = \frac{2}{g}\,\mathcal{H} \,\hat\lambda \,.
\label{cuant}
\en
The above choice for $B_i$ is consistent with the following embeddings of translations:
\be
\label{twist-suN}
T_i(y) = e^{\epsilon_{ij} \pi i(k+\frac{m}{N}) \,\frac{y_j}{l_j}  \,\hat\lambda} \, ,
\en
which satisfy the conditions in Eq.~(\ref{conmut}), when $\hat\lambda$ is chosen as the 
$SU(N)$ generator of the Cartan subalgebra of the form $\hat{\lambda}= diag(1,1,\cdots, 1-N)$.

The boundary conditions for the fluctuation fields can be most conveniently expressed choosing
the bases in Poincar\'e space defined by $z(\overline{z})\equiv (y_1 \pm i y_2)/\sqrt{2} $
and ${A}_{z(\overline{z})}^a\equiv (A^a_1 \mp i A^a_2)/\sqrt{2} $
and in gauge space by $[\lambda_a,\hat\lambda]= q^a \lambda^a\,$.
In these bases,
\be
\left\{
\begin{array}{ccc} 
A_{M}^{a} (y_1 + l_1, y_2) &=&  e^{i\,\pi (k+\frac{m}{N}) \frac{y_2}{l_2}\, q^a} \;A_{M}^{a} (y_1, y_2) \\
A_{M}^{a} (y_1, y_2 + l_2) &=&  e^{-i\,\pi (k+\frac{m}{N}) \frac{y_1}{l_1}\, q^a} \;A_{M}^{a} (y_1, y_2) \,,
\end{array} \right.
\label{periooo1}
\en
\be
D_z^a = \partial_{z} - \frac{\mathcal{H}}{2}\, \overline{z}\,q^a\,\hspace{0.5cm} , \hspace{0.5 cm}
D_{\bar{z}}^a = \partial_{\overline{z}} + \frac{\mathcal{H}}{2}\, z \,q^a
\quad {\rm with} \quad
\left[ D_z^a , D_{\bar{z}}^a \right] = \, {\mathcal H}\, q^a\,.
\en
The non-commutativity of the {\it imposed}-background covariant derivatives, acting on charged fields, illustrates that translations of arbitrary length along the two extra dimensions do not commute.
In order to determine the physical spectrum, all terms in the Lagrangian in Eqs..~(\ref{L0})-(\ref{L4})
will have to be considered.

\subsubsection*{Total background}

Were the Lagrangian  formally expanded instead around an hypothetical {\it total} minimum 
with background $\widetilde B_M(y)$, Eq.~(\ref{absin}), and its fluctuations\footnote{$A^a_M$ 
is used throughout the paper to generically denote excitations with respect to the background 
included in any definition of the covariant derivative.}, the corresponding $\widetilde G_{MN}$ 
would vanish,
\be
{\widetilde{G}}_{MN}=\frac{i}{g}[\widetilde{D}_M,\widetilde{D}_N]=0\,,
\label{conm12total}
\en
with $\widetilde{D}_M$ given by
  \be
{\widetilde D}_M A_N^a \equiv \partial_M A_{N}^a  - g  f^{abc} A_N^{b} {\widetilde B}_M^c \,.
\label{totalderivataback}
\en
No tachyonic mass would be present then in the Lagrangian and, to extract the physical spectrum, it would be enough
to consider only terms with two fluctuation fields,
 \be
\mathcal{\widetilde L}^{(2)}_{A}\equiv -\frac{1}{2} [\widetilde D_M A_N^{a} \,\widetilde D^M A^{N\,a} -\widetilde D_M A_N^{a} \,\widetilde D^N A^{M\,a}]\, .
\en
Below we will explicitly explore the dynamical evolution of the system from the {\it imposed}
background $B_M( y)$ to the {\it total} stable one, $\widetilde B_M( y)$, in the $SU(2)$ case.

\subsection{The $6$-dimensional $SU(2)$ Lagrangian}

We particularize now the discussion to a $SU(2)$ gauge theory, with generators 
$\lambda^a = \sigma^a/2$, where $a=1,2,3$ and  $\sigma^a$ denote the Pauli matrices.
The commutativity condition for the embeddings of translations in gauge space, Eq.~(\ref{conmut}),
reduces now to the values $\pm1$, as $m$ can take only two values, $m=0,1$, while $k$ keeps being an arbitrary integer. A possible choice for the {\it imposed} background is  one pointing towards
the third gauge direction, i.e. $\hat\lambda=\sigma_3/2$, whose replacement in
Eqs.~(\ref{back-suN}-\ref{periooo1}), defines the background and boundary conditions for this case.
The gauge indices for fields in the adjoint representation are $a\,=\,+,-,3$, with
\be
\label{cambio-adjoint}
\left\{
\begin{array}{c}
\lambda^+ = \frac{1}{\sqrt{2}} (\lambda_1 + i \lambda_2) \\
\lambda^- = \frac{1}{\sqrt{2}} (\lambda_1 - i \lambda_2)
\end{array}
\right.
\hspace{2em} {\rm and} \hspace{2em}
\left\{
\begin{array}{c}
\mathbf{A}^+_M = \frac{1}{\sqrt{2}} (\mathbf{A}^1_M - i \mathbf{A}^2_M) \\
\mathbf{A}^-_M = \frac{1}{\sqrt{2}} (\mathbf{A}^1_M + i \mathbf{A}^2_M)
\end{array} \,\,\,\,,
\right.
\en
where $M= \mu, z,\overline{z}$. For those fields, the charges with respect to the 
{\it imposed} background are $q^a=\,+2,-2,0$, in units of the charge of the fundamental 
representation, $q_f=1/2$.

Consider the various components of the Yang-Mills Lagrangian,  Eqs.~(\ref{L0})-(\ref{L4}), for
the particular case of $SU(2)$. Working in the basis of Eq.(\ref{cambio-adjoint}),
the Lagrangian without gauge fixing terms can now be explicitly expanded as
\be
\mathcal{L}_{6D} = \mathcal{L}_{\mu\nu} + \mathcal{L}_{ij} + \mathcal{L}_{\mu\,i}\,,
\en
where
\be
\mathcal{L}_{\mu\nu} &=& -\frac{1}{4} F^a_{\mu \nu} F^{\mu \nu}_a
\label{gaugelag} \\ \nn \\
\mathcal{L}_{i j} &=& 2\,{\mathcal H}  \left( A_{\overline{z}}^- A_z^+-A_{\overline{z}}^+ A_z^-\right)+\frac{1}{2} \left[
 (\partial_{\overline{z}} A_z^3)^2  +  (\partial_{z} A_{\overline{z}}^3)^2  - 2\, (\partial_{z} A_{\overline{z}}^3) (\partial_{\overline{z}} A_{z}^3)  \right]  \label{scalpot} \\
 &+& \left[(D_{\overline{z}} A_z^+) (D_{\overline{z}} A_z^-) +  
           (D_{z} A_{\overline{z}}^+) (D_{z} A_{\overline{z}}^-) -  
           (D_{z} A_{\overline{z}}^+) (D_{\overline{z}} A_z^-)   -
           (D_{\overline{z}} A_{z}^+) (D_{z} A_{\overline{z}}^-) \right] \nonumber \\
&-&  g^2 \left[\frac{1}{2} (A_z^+ A_{\overline{z}}^- -A_{\overline{z}}^+ A_z^-)^2
+ A_{z}^3 A^3_{\bar{z}} \left( A_z^+ A_{\overline{z}}^- + A_z^- A_{\overline{z}}^+\right)\right] \nn \\
&-& g^2\left[A_{z}^3A_{z}^3  A_{\overline{z}}^+ A_{\overline{z}}^- + \rm{h.c.} \right]
+ i  g  \left(A_z^+ A_{\overline{z}}^- - A_{\overline{z}}^+ A_z^-\right)
\left( D_{\bar{z}} A_{z}^3 - D_z  A^3_{\bar{z}}\right)\nonumber\\
&+& i g \left[\left(A_z^3 A_{\overline{z}}^+ - A^3_{\bar{z}} A_z^+\right)
 \left(D_{\bar{z}} A_z^- - D_z A_{\overline{z}}^-\right)- \rm{h.c.}\right]\,,\nn \nn \\ \nn \\
\mathcal{L}_{\mu i}&=& g^2
 ( A_{\mu}^+ A^{\mu}_-(2 A^3_{\bar{z}} A^3_z + A_z^+A_{\overline{z}}^- +A_{\overline{z}}^+A_z^-) +
A_\mu^3 A^\mu_3( A_z^+A_{\overline{z}}^- +A_{\overline{z}}^+A_z^-)  \label{mixpot} \\
&-& \left[ A_\mu^3 A^\mu_+(A_z^3 A_{\overline{z}}^- + A_{\bar{z}}^3 A_z^-) + h.c.\right]
-\left[  A^\mu_+ A_\mu^+ A_{\overline{z}}^- A_z^- + \rm{h.c.} \right] ) \nonumber \\
&+& i g [ (\partial_\mu A_z^3 - D_z A_\mu^3)(A^\mu_- A_{\overline{z}}^+ - A^\mu_+ A_{\overline{z}}^-) +
(\partial_\mu A_z^+ - D_z A_\mu^+)(A^\mu_3 A_{\overline{z}}^- - {A_{\bar{z}}}^3 A^\mu_-) \nonumber \\
&+& (\partial_\mu A_z^- - D_z A_\mu^-)(A^\mu_+ {A_{\bar{z}}}^3 - A^\mu_3 A_{\overline{z}}^+) -\rm{h.c.} ]
\nn \\
&+& \partial_\mu A_\mu^+ \,(D_z A_{\overline{z}}^- + D_{\bar{z}} A_z^-)+
\partial_\mu A_\mu^-\, (D_z A_{\overline{z}}^+ + D_{\bar{z}} A_z^+)+
 \partial_\mu A_\mu^3\,(D_z A_{\bar{z}}^3 + D_{\bar{z}} A_z^3)\nonumber\,.
\en
From the $4D$ point of view, $\mathcal{L}_{\mu \nu}$, $\mathcal{L}_{i j}$ and $\mathcal{L}_{\mu i}$ 
will generate - after fixing the gauge - the pure gauge Lagrangian, the scalar potential and 
the gauge invariant kinetic terms of the scalar sector, respectively.  Notice the term 
$2 {\mathcal H}  \, A_{\overline{z}}^- A_z^+$ in  $\mathcal{L}_{ ij}\,$: it corresponds 
to a negative mass squared for the $A_z^+$ field, which pinpoints the instability of the 
theory expanded around a false vacuum.

\subsubsection*{Gauge fixing Lagrangian: the $R_\xi^{6D}$ gauge}
The structure of the $\mathcal{L}_{\mu i}$ term suggests immediately a certain gauge choice
compatible with the boundary conditions, that we will call the $R_\xi^{6D}$ gauge. Among all terms
in the $6D$ Lagrangian containing two fluctuation fields, i.e. $\mathcal{L}^{(2)}_{A}$, the only $4D$ derivative
 interaction of  the $A_{\mu}$  is of the form
\be
- A^{\mu}_a \partial_{\mu} \left( D_z A_{\overline{z}}^a + D_{\overline{z}} A_z^a\right) \,,
\label{su2gf}
\en
and it appears in the last row of $\mathcal{L}_{\mu i}$. These terms are cancelled by the following
choice for the gauge-fixing Lagrangian
\be
\mathcal{L}^{g.f.}_{6\xi} = -\frac{1}{2\,\xi} \sum_a\,\left[ \partial_{\mu} A^{\mu}_a \,
- \,\xi\, \left({D}_z A_{\bar{z}}^a\,+\, {D}_{\bar{z}} A_z^a  \right) \right]^2 \,.
\label{gaugefix}
\en
A warning is pertinent here. Not all terms which lead to $4D$ mixed terms (bilinears involving $4D$ derivatives of gauge fields and  scalar fields) will be eliminated through this gauge choice. Additional $4D$ 
mixed terms may result from the cubic couplings appearing in the third and fourth rows of 
$\mathcal{L}_{\mu i}$, if some $4D$ scalars take vacuum expectation values due to the 
instability of the present expansion of the Lagrangian. In other words, the naive 
$R_\xi^{6D}$ gauge defined above does not match a proper $4D$  $R_\xi$ gauge.
We will come back to this point later on, in subsection~3.4.

\subsection{The effective $4$-dimensional $SU(2)$ Lagrangian}

The $4D$ Lagrangian,
\be
\mathcal{L}^{4D}= \int_{\mathcal{T}_2} \hspace{-0.25cm } d^2 y\,\mathcal{L}(x,y)\,,
\en
 will describe the physics of $4D$ fields, $A_M^{a\,(r)}(x)$, defined from
\be
A_M^a (x,y)\equiv \sum_r^{} A_M^{a\,(r)}(x) f^{a(r)}(y) \,,
\en
with the extra-dimensional wave functions $f^{a(r)}$ satisfying  the boundary conditions
\be
\left\{
\begin{array}{ccc}
f^{a(r)} (y_1 + l_1, y_2) &=& e^{  i \pi  (k + \frac{m}{N}) \frac{y_2}{l_2}\, q^a} \;f^{a(r)} (y_1, y_2)\,, \\
f^{a(r)} (y_1 , y_2 + l_2) &=&  e^{ - i \pi (k + \frac{m}{N}) \frac{y_1}{l_1}\, q^a} \;f^{a(r)} (y_1, y_2)\,,
\label{periooo1-bis}
\end{array} \right.
\en
and $r$ referring to the infinite towers of $4D$ modes. Depending on their gauge charge, fields are
neutral ($a=3$) or charged ($a=\pm$) with respect to the {\it imposed} background, and may be
arranged in $4D$ KK towers ($r=n_1,n_2$) for the former and Landau levels ($r=j$)
for the latter.

The shape of the extra-dimensional wave functions depends exclusively on the boundary conditions,
encoded in the covariant derivative. That is, the wave functions depend on the gauge index
(whether neutral or charged with respect to the background), but do not depend on its Lorentz
index (whether $4D$ vectors or scalars).

\subsubsection*{Neutral fields}
For neutral fields, the covariant derivatives $D_i$ reduce to ordinary (commuting) derivatives.
For the $4D$ vectors $A^{3\,(n_1,n_2)}_\mu(x)$, the following masses result 
\be
(\partial_z \partial_{\overline{z}} + \partial_{\overline{z}} \partial_z) f^{3 \,(n_1,n_2)}(y)
&=& m^2_{3\,(n_1,n_2)}  f^{3 \,(n_1,n_2)}(y) \,,
\label{moto-neutro}
 \en
where
\be
m^2_{3\,(n_1,n_2)}\equiv  \,4 \pi^2 \,\left(\frac{n_1^2}{l_1^2} + \frac{n_2^2}{l_2^2}\right)  \,,
\label{neutralm}
\en
while the eigenfunctions are given by
\be
f^{3(n_1,n_2)}(y)= \frac{1}{\sqrt{\mathcal{A}}} \,e^{2 \pi i\,\left(n_1 \frac{y_1}{l_1}+ n_2 \frac{y_2}{l_2}\right)}\,.
\en
The mode $A^{3\,(0,0)}_\mu(x)$ remains massless at this level, as it would for a residual $U(1)$
symmetry.

For neutral scalar fields, the quadratic mass terms in the $R_\xi^{6D}$ gauge, 
Eqs.~(\ref{scalpot}) and (\ref{gaugefix}), lead to the following $4D$ Lagrangian 
after integration over the extra dimensions,
 \be
\left(\mathcal{L}^{4D}_{ij}\right)^{neutral}_2 \hspace{-0.3cm}
        = -\frac{1}{2} \hspace{-0.3cm} \sum_{n_1,n_2=-\infty}^{\infty} \hspace{-0.3cm} m_{3\,(n_1,n_2)}^2
           \left\{A^{(-n_1,-n_2)}(x)\,A^{(n_1,n_2)}(x) \,+ \,
           \xi \,a^{(-n_1,-n_2)}(x)\,a^{(n_1,n_2)}(x)\,\right\}\,, \nn
\label{L4Dij}
\en
where $A^{(n_1,n_2)}(x)$ and $a^{(n_1,n_2)}(x)$ are the mass eigenstates,
\be
 a^{(n_1,n_2)}(x) &\equiv& \frac{-i}{\sqrt{2}} \left(e^{i\theta_{(n_1,n_2)}} A_z^{3\,(n_1,n_2)}(x)
 + e^{-i \theta_{(n_1,n_2)}} A_{\overline{z}}^{3\,(-n_1,-n_2)}(x) \right) \,, \\
 A^{(n_1,n_2)}(x) &\equiv& \frac{1}{\sqrt{2}} \left(e^{-i\theta_{(n_1,n_2)}} A_{\bar{z}}^{3\,(-n_1,-n_2)}(x)
 - e^{i \theta_{(n_1,n_2)}} A_{z}^{3\,(n_1,n_2)} (x)\right)\,,
\label{aA}
\en
with $e^{i \theta_{(n_1,n_2)}} \equiv \frac{2\pi}{m_{3(n_1,n_2)}}\, \left(\frac{n_1}{l_1}+ i \frac{n_2}{l_2}\right)\,$.

In the absence of instability, $A^{(n_1,n_2)}(x)$ would be the physical neutral scalar fields,
while $a^{(n_1,n_2)}(x)$ would play the role of pseudo-Goldstone bosons, eaten by the
$A_\mu^{3\,(n_1,n_2)}(x)$ to acquire mass. Notice that indeed the quantity
$D_z A^3_{\bar{z}} + D_{\bar{z}} A_z^3$ appearing in the gauge fixing condition, Eq.~(\ref{gaugefix}),
can be expressed in terms of the  scalars $a^{(n_1,n_2)}$ alone:
\be
D_z A^3_{\bar{z}}+ D_{\bar{z}} A_z^3 =  - \sum_{n_1,n_2=-\infty}^{\infty}
m_{3(n_1,n_2)}\,a^{(n_1,n_2)}(x)\, f^{(n_1,n_2)} (y) \,.
\en
Notice as well that it does not exist a pseudo-Goldstone boson with $n_1=n_2=0$, 
which is consistent with the fact that $A_{\mu}^{3\,(0,0)}$ has not received, 
at this level, a contribution to its mass.

\subsubsection*{Charged fields}
To determine the Landau energy levels, define as usual
creation and destruction operators $a$ and $a^{\dagger}$, for charges $q^{\pm}=\pm2$,
\be
\begin{array}{lcr}
a_{+} \,\equiv \,-\frac{ i}{\sqrt{2 {\mathcal H}}}\,D_{\overline{z}}^{(+)}\,,
& & a_{-} \,\equiv \,\frac{ i}{\sqrt{2 {\mathcal H}}}\,D_{z}^{(-)}\,, \\
a^\dag_{+} \,\equiv \,- \frac{ i}{\sqrt{2 {\mathcal H}}}\,D_z^{(+)}\,,
& & a^\dag_{-} \,\equiv \,\frac{ i}{\sqrt{2 {\mathcal H}}}\,D_{\overline{z}}^{(-)}\,,
\end{array}
\label{def-ope}
\en
which satisfy commutation relations
\be
\left[ a_{\pm} , a^\dag_{\pm} \right] = 1 \;.
\en
Defining as well the number operator $\hat{j}_{(\pm)} = a_{(\pm)}^\dagger a_{(\pm)}$, 
it results that charged fields  
 $A_{M}^{\pm\,(j)} (x)$ get at least partial contributions to their masses from the term
\be
-(D_z^{a} D_{\bar{z}}^{a} + D_{\bar{z}}^{a} D_z^{a})\,  f^{a(j)}(y) = m^2_{a\,(j)}\,  f^{a(j)}(y) \,,
\label{masseSU(2)}
\en
with $a=\pm$ and mass eigenvalues given by
\be
m^2_{\pm\,(j)} \equiv 2 {\mathcal H} (2 j + 1)\,=\,\frac{4 \pi(k+\frac{m}{2})}{\mathcal{A}}\,(2j+1)
\,,
\label{chargedm}
\en
where $j$ integer $\ge 0$.

That is, for charged fields the commutator in Eq.~(\ref{conm12}) does not vanish and in 
consequence no zero eigenvalues are expected. In other words, while neutral fields can be 
simultaneously at rest with respect to the two extra dimensions, charged fields cannot, 
as  a charged particle in a magnetic field moves. The energy levels are  Landau levels.
Notice as well that the mass scale is set by the torus area, the 't Hooft non-abelian flux 
$m$ and the integer $k$, while it is independent of the $6D$ coupling constant $g$.

The associated extra-dimensional wave functions,
\be
 {f^{+}}^{(j,\rho)} (x,y) &=&
\left( \frac{2 \,d}{l_1^3 l_2} \right)^{\frac{1}{4}} \, \frac{(-i)^j}{\sqrt{2^j\,j!}}\,
e^{i \pi \,d \,\frac{y_1 y_2}{l_1\,l_2}} \quad \times \label{pesados-no-ab} \\
                      & \, & \hspace{-1.5cm}
\sum_{n=-\infty}^{\infty}  e^{- \frac{ \pi d}{l_1 l_2} (y_2 + n l_2 + \frac{\rho l_2}{ d})^2}
                          \, e^{2 \pi i (d\, n + \rho) \frac{y_1}{l_1} } H_{j,\rho} 
         \left[\sqrt{\frac{2 \pi d}{l_1\,l_2}} \left( y_2 + n l_2 + \frac{\rho l_2}{ d} \right)\right] \nn 
\en
are derived explicitly in Appendix \ref{Landau_Levels}. The opposite-charge field is ${f^{-}}^{(j,\rho)}(x,y)\,=\,\left({f^{+}}^{(j,\rho)}(x,y)\right)^*$.
Obviously,  ${f^{+}}^{(j,\rho)}$ and ${f^{-}}^{(j,\rho)}$ satisfy the boundary conditions 
in Eq.~(\ref{periooo1-bis}).

The quantity $d$ in Eq.~(\ref{pesados-no-ab}) is defined by
\be
d \equiv q\,(k+\frac{m}{N})\,,
\label{degeneracy}
\en 
and signals degeneracy. Notice the index $\rho$: generically, the tower of Landau levels may be defined by 
another quantum number \cite{Giusti:2001ta} in addition to $j$. $\rho$ sweeps over 
these extra degrees of freedom,
 \be 
 0\,\le\,\rho\,\le\,d-1\,, 
 \label{rho}
 \en
and its possible values signal degenerate energy levels, as the latter are independent 
of $\rho$, see Eq.~(\ref{chargedm}) above. For a field of given charge $q$ (i.e, $q=2$ 
and $q=1$ for fields in the adjoint and fundamental representation of $SU(2)$, respectively), 
the degree of degeneracy is given by $d$. As discussed in Appendix~\ref{Landau_Levels}, $d$ is necessarily an integer, which for $SU(2)$ reduces to 
either $d=qk$ or $d=q(k+\frac{1}{2})$, depending on the value of $m$.

While $4D$ charged vectors $A_\mu^{\pm(j,\rho)}$ get only mass contributions from  
Eq.~(\ref{chargedm}) above, charged scalars receive further contributions from quadratic 
terms in Eq.~(\ref{scalpot}). Working in the $R_\xi^{6D}$ gauge, Eq.~(\ref{gaugefix}), 
and, after diagonalizing the system, we obtain
\be
\left({\mathcal{L}}^{4D}_{ij}\right)_2^{charged} &=& \sum_{\rho=0}^{d-1} \, 
 \Bigg\{\left. 2 {\mathcal H} \, {H}^*_{0,\rho}(x) {H}_{0,\rho}(x)
 - 2 {\mathcal H} \, \sum_{j=1}^{\infty} (2j+1) H_{j,\rho}^*(x)\,H_{j,\rho}(x) \right. \nn \\
& & \hspace{1.0cm} -\xi\, 2 {\mathcal H}\,\sum_{j=0}^{\infty} (2j+1) h_{j,\rho}^*(x)\,h_{j,\rho} (x)
    \Bigg\} \,.
\en

This Lagrangian has been written in terms of the following mass eigenfunctions:
\be
& & H_{0,\rho}(x) = - A_{\overline{z}}^{-\,(0,\rho)}(x)\,, \nn \\
& & h_{0,\rho}(x)= A_{\overline{z}}^{-\,(1,\rho)}(x)\,,\nn \\
& & H_{j,\rho}(x) = 
           - s_j A_{\overline{z}}^{-\,(j+1,\rho)} (x) + c_j A_z^{-\,(j-1,\rho)}(x) \,, \nn \\
& & h_{j,\rho}(x) = c_j A_{\overline{z}}^{-\,(j+1,\rho)}(x) + s_j A_z^{-\,(j-1,\rho)}(x)\,,
\label{xieigen}
\en
where $c_j \equiv \cos \theta_j = \sqrt{\frac{j+1}{2 j +1} }$ and $s_j \equiv \sin \theta_j = \sqrt{\frac{j}{2 j +1}}$, with $j\ge 1$. $H_{0,\rho}(x)$ denotes the $4D$ field (or fields, when $\rho$ takes several values)
 with negative mass(es) 
$-2{\mathcal H}$ and $h_{0,\rho}(x)$ its unphysical scalar partner(s), eaten -at this level- by the $A_{\mu}^{+\,(0,\rho)}(x)$ field(s) to become massive\footnote{The tachyon $H_{0,\rho}$ could 
also be correctly denoted $H_{-1,\rho}$, as a $j=-1$ state, extending the definition given 
for the $H_{j,\rho}$ fields. We have refrained from doing so, though, with the aim of beautifying the notation.}.

In the absence of the instability induced by the negative mass,
$H_{j,\rho}(x)$ would be the physical charged scalar fields, while $h_{j,\rho}(x)$ would play
the role of pseudo-Goldstone bosons, eaten by the $A_{\mu}^{+\,(j,\rho)}(x)$ fields to acquire mass.
Indeed, the gauge fixing condition can be expanded as
\be
D_z A_{\overline{z}}^-  + D_{\bar{z}} A_z^- = i \sum_{\rho=0}^{d-1} \sum_{j=1}^{\infty}  \, m_{\pm j} \, h_{j,\rho}(x) \, f^{-(j,\rho)}(y)\,.
\label{www2}
\en
Notice as well that this result holds for any value of $j$, including  $j=0$,
since ${A_{\mu}}^{\pm\,(0,\rho)}(x)$  has taken a contribution to its mass after compactification,
 as a consequence of its interaction with the imposed background.

The Lagrangian exhibits thus a behavior  that could correspond to the breaking 
$SU(2)\rightarrow U(1)$, although the presence of the tachyon  $H_{0,\rho}(x)$ 
signals that the true vacuum remains to be found.  The remaining analysis can be
technically simplified working in the $R^{6D}_\xi$ gauge with $\xi=\infty$: the 
would-be goldstone fields $a(x)$ and $h(x)$ disappear then from the analysis, and 
results will be given for this case. However, before proceeding to it, let us 
briefly discuss another gauge-fixing choice, alternative to that used above.

\subsection{The $R_\xi^{4D}$ gauge}
\label{expansiontrue}

An appropriate gauge choice, also compatible with the boundary conditions, is
 \be
\mathcal{L}^{g.f.}_{4\xi} = -\frac{1}{2\,\xi} \sum_a\,\left[ \partial_{\mu} A^{\mu}_a \,
- \,\xi\, \left({\widetilde D}_z A_{\bar{z}}^a\,+\, {\widetilde D}_{\bar{z}} A_z^a  \right) \right]^2 \,,
\label{gaugefix4}
\en
where now ${\widetilde D_i}$ is the {\it total} covariant derivative defined in Eq.~(\ref{totalderivataback}), corresponding to a stable background. Notice the analogy 
with the analysis in the previous subsections in terms of the $R_\xi^{6D}$ gauge, 
Eq.~(\ref{gaugefix}). The choice in Eq.~(\ref{gaugefix4}) guarantees the elimination 
of {\bf all} $4D$ scalar-gauge crossed terms stemming from the  last three rows of 
$\mathcal{L}_{\mu i}$, Eq.~(\ref{mixpot}), including those resulting after spontaneous 
symmetry breaking. It is  then a true $R_\xi$ gauge from the four-dimensional point of view.

In this gauge, it is trivial to formally identify the terms in the 6D Lagrangian which will 
give rise to the masses of the different type of $4D$ fields: gauge bosons and their replica, 
physical scalars and ``would be'' goldstone bosons:

\begin{enumerate}
\item Gauge boson masses will result from
\be
\mathcal{L}_{mass}^{gauge} = - \frac{1}{2} \,A_{\mu}^a \,\left[ {\widetilde D}_z\,{\widetilde D}_{\bar{z}}+\,{\widetilde D}_{\bar{z}}\,{\widetilde D}_z \,\right]_{ab} \, A^{\mu\,b}\,,
\label{4mgauge}
\en
where $a,b$ are the indices in the adjoint representation.
\item Physical, $\xi$-independent, scalar masses will stem from
\be
\mathcal{L}_{mass}^{scal} &=& - \frac{1}{2} \,\left( {\widetilde D}_z A_{\bar{z}}^a\,-\,{\widetilde D}_{\bar{z}} A_z^a \right)^2 \nonumber \\
&=& - \frac{1}{2}\,\left(
A_z^a ,\,A_{\bar{z}}^a
\right)\,
\left(
\begin{array}{cc}
- {\widetilde D}_{\bar{z}} {\widetilde D}_{\bar{z}} & {\widetilde D}_{\bar{z}} {\widetilde D}_z \\
{\widetilde D}_z {\widetilde D}_{\bar{z}} & - {\widetilde D}_z {\widetilde D}_z
\end{array}
\right)_{ab}\,\left(
\begin{array}{c}
A_z^b \\
A_{\bar{z}}^b
\end{array}
\right) \,.
\label{4mphyscalar}
\en
Because $[{\widetilde D}_z, {\widetilde D}_{\bar{z}}]=0$ (see Eq.~(\ref{conm12total})), 
the eigenvalues of this matricial equation produce the following mass contributions 
to scalar fields:
\be
\begin{array}{ccc}
\Delta M^2_{physical}  & = & \frac{1}{2} \,\left[{\widetilde D}_z {\widetilde D}_{\bar{z}}\,+\, 
{\widetilde D}_{\bar{z}} {\widetilde D}_z  \right] \,,\\
\Delta M^2_{goldstone} & = & 0\,.
\end{array}
\label{4mscalar}
\en
Comparison with Eq.~(\ref{4mgauge}) shows that  it is  generically expected to find a scalar 
partner for each 4D gauge boson, degenerate in mass.

\item Finally, the $\xi$-dependent scalar masses will result from,
\be
\mathcal{L}_{mass}^{\xi} &=& - \frac{\xi}{2} \,\left( {\widetilde D}_z A_{\bar{z}}^a\,+\,{\widetilde D}_{\bar{z}} A_z^a \right)^2 \nonumber \\
&=&  \frac{1}{2}\,\left(
A_z^a, \,
A_{\bar{z}}^a
\right)\,\left(
\begin{array}{cc}
 {\widetilde D}_{\bar{z}} {\widetilde D}_{\bar{z}} & {\widetilde D}_{\bar{z}} {\widetilde D}_z \\
 {\widetilde D}_z {\widetilde D}_{\bar{z}} &  {\widetilde D}_z {\widetilde D}_z
\end{array}
\right)_{ab}\,\left(
\begin{array}{c}
A_z^b \\
A_{\bar{z}}^b
\end{array}
\right) \,.
\label{4munphyscalar}
\en
Once again, because ${\widetilde D}_z$ and $ {\widetilde D}_{\bar{z}}$ commute, the eigenvalues of $\mathcal{L}_{mass}^{\xi}$ will result in mass contributions
\be
\Delta M^2_{goldstone} &=& \frac{\xi}{2} \,\left[{\widetilde D}_z {\widetilde D}_{\bar{z}}\,+\, {\widetilde D}_{\bar{z}} {\widetilde D}_z  \right]\,, \nonumber \\
\Delta M^2_{physical} &=& 0\,.
\label{4mgoldstone}
\en

\end{enumerate}

The coincidence between the eigenvalues expected for the gauge and ``would be'' goldstone 
boson masses is a characteristic of hidden non-abelian symmetries. The larger degeneracy 
among the three sectors -gauge, physical scalars and unphysical scalars- is related to the 
fact that {\it total} field strength of the stable vacuum is zero. In consequence the coordinate dependent 
conditions are equivalent to constant ones, as shown in Section 2, which discriminate 
among gauge charges, not among Lorentz indices.

In the next Section, we will follow the dynamical evolution of the system towards a stable 
vacuum, determining  the minimum of the $4D$ potential and obtaining the physical spectra 
in both the $R_\xi^{4D}$ and $R_\xi^{6D}$ gauges.

\section{The minimum of the $4$-dimensional potential}
Below, we will obtain the effective $4D$ potential for $SU(2)$, minimize it and 
find the physical spectra. The results will be compared with the theoretical 
expectations developed in Section 2.

We have first integrated the $6D$ Lagrangian,  Eqs.~(\ref{gaugelag})-(\ref{mixpot}), plus 
the gauge-fixing term, Eq.~(\ref{gaugefix}) or Eq.~(\ref{gaugefix4}), over the $2D$ torus 
surface, obtaining in this way all effective $4D$ couplings among the towers of states. 
In ordinary compactifications, i.e. without background with constant field strength,
a good understanding of the $4D$ light spectrum only requires to consider the lightest KK 
states and their self-interactions. With the inclusion of such background, this is no more 
the case due to the simultaneous presence of KK and Landau levels. Cubic and quartic terms 
link a given neutral (KK) field to an infinity of charged (Landau) levels, and viceversa. 
Previous analysis of scenarios with background with constant field strength, 
such as the original Nielsen and  Olesen one~\cite{Nielsen}, as well as subsequent 
studies \cite{Antoniadis}, have typically 
included only quartic interactions of the lowest $4D$ charged level (i.e. the tachyon), 
with at most the addition of the tower of only one type of replica. However, we will 
show that it is necessary to consider many modes and all types of interaction between 
KK and Landau levels, for a true understanding of the system.

For quadratic terms, the integration over the torus reduces to the use of the orthogonality 
relations for the bases of extra-dimensional wave functions. The inclusion of cubic and quartic 
interactions requires the evaluation of integrals with three and four extra-dimensional wave 
functions. We have solved them analytically in the general case. The results can be
found in Appendix \ref{AppIntegrals}, together with the completeness relationships linking 
them. The latter have been checked as well numerically up to a precision better than $10^{-6}$.

We have then proceeded to look for the minima of the potential. Let us previously recall 
the theoretical expectations. As  the true vacuum should have total zero energy, see 
Eq.~(\ref{truevacuum}), the stable minimum of the $SU(2)$ $4D$ potential should correspond
to a dynamical reaction of the system of the form
\be
 F_{12}^3(x,y)\big|_{min} = -G_{12}^3 \,=\, \frac{2 \mathcal{H}}{g}
                          \,=\, \frac{4 \pi}{g\,\mathcal{A}} (k+\frac{m}{2})\,,
\en
so as to cancel the contribution of the {\it imposed} background. 
That is, the following value for the minimum of the $4D$ potential is expected (see Eq.(\ref{vmin4D})):
\be
 V\big|_{min} = \frac{1}{2} \int_{\mathcal{T}^2} \, d y \, 
  [(F_{12}^3(x,y))^2 + 2 \, G_{12}^3\,F_{12}^3(x,y)]\
 \big|_{min}
            \,=\, - \frac{8\pi^2}{g^2 \mathcal{A}}\, (k+\frac{m}{2})^2\,.
\label{vminvalue}
\en
We analyze below whether the minimum of our $4D$ effective potential does converge towards such values.
Three comments on the procedure are pertinent:
  \begin{enumerate}
\item The determination of the set of vacuum expectation values that 
      minimizes the potential can only be done  numerically. Starting with the inclusion of only the 
      lightest fields of the KK and Landau towers,  heavier replicas of both types will be successively 
      added and the corresponding minimum determined at each step. The total number of neutral and charged 
      replica to be included in the analysis is determined requiring that the minimum of the potential 
      reaches an asymptotically stable regime.
\item For technical and theoretical reasons, we will present our results in the two gauges previously 
      described: the $R_\xi^{6D}$ gauge, for the particular case $\xi=\infty$, and the general 
      $R_\xi^{4D}$ gauge. 
      This will allow precise checks of the gauge invariance of the results. 
\item In order to keep as low as possible the degeneracy of states, while analyzing the two possible 
      non-trivial setups, the numerical results will be confined to two cases: a) $m=0\,,\, k=1$ and 
      b) $m=1\,,\,k=0$.   Furthermore, all numerical results presented below correspond to an 
      isotropic torus\footnote{ The anisotropic case will be considered in a future work.}, $l_1=l_2$.
  \end{enumerate}

\subsection{Non-trivial 't Hooft flux: $m=1$, $k=0$}

This case corresponds to a non-trivial 't Hooft flux, in which the generators of the 
translation operators $T_i$ anti-commute. The fields in the Landau towers are not degenerate, 
as $d=1$ in Eq.~(\ref{degeneracy}): the index $\rho$ become thus meaningless and it will be 
obviated all through this Subsection.

Let us illustrate with a simple argument how the system dynamically approaches the true vacuum 
and the need of including rather high neutral and charged modes. Consider for the 
moment only the charged scalar zero mode, $H_{0}$ (i.e. the tachyon), the lightest neutral 
scalar $A^{3\,(0,0)}_z$ and their interactions. The effective $4D$ potential is then simply 
given by:
\be V &=& - 2
\mathcal{H} \,\,|H_{0}(x)|^2  + \frac{g^2}{2}\,\,I^{(4)}_0 \,\,
|H_{0}(x)|^4 + |H_{0}(x)|^2\, A^{3\,(0,0)}_z(x)\,A_{\bar{z}}^{3\,(0,0)}(x)\,,
\label{zero-}
\en with $I^{(4)}_0$
referring to the 4-point integral between the lightest charged states\footnote{The general 
definition of the 3-point and 4-point integrals is given in Appendix~\ref{AppIntegrals}. Here 
$I^{(4)}_0$ is an abbreviated notation for the integral $I^{(4)}_C[0,0,0,0,0,0,0,0]$ defined there.}.
One can immediately recognize in Eq.~(\ref{zero-}) the classical mexican-hat potential,  with
its minimum corresponding to:
\be
 < |H_{0}(x)|^2 > &=& \frac{2 \mathcal{H}}{g^2 I^{(4)}_0} \,, \qquad
 < A^{3\,(0,0)}_z (x) > = < A^{3\,(0,0)}_{\bar{z}} (x) > \ = \ 0 \,.
\label{minimo-tac}
\en
In this simplified example, only the charged scalar (i.e.the tachyon) acquires a non vanishing 
vacuum expectation value 
({\it vev}) while the neutral fields remain unshifted. Using the numerical value $1/I^{(4)}_0 = 
(0.85 \, \mathcal{A})$,
it results\footnote{The dimensions of the quantities in Eq.~(\ref{minimo-tac}) are  
$[\mathcal{H}] = [I^{(4)}_0] = [E^2]$ and $[g]=[E^{-1}]$.}:
 \be
 V_{min}= - \frac{2 \mathcal{H}^2}{g^2 I^{(4)}_0}
         \sim -0.85 \times \frac{2 \pi^2}{g^2\mathcal{A}} \, ,
 \en
which is still quite different from that predicted by  
 Eq.~(\ref{vminvalue}). Moreover, it is enough to add the interactions with either 
the next neutral or charged levels to observe the appearance of tadpole terms. That is, 
the true minimum of the system does not correspond then anymore to the {\em vev}s 
obtained in Eq.~(\ref{minimo-tac}), but all fields get new shifts instead. 
\FIGURE[t]{
\vspace{-0.5cm}
\includegraphics[width=13cm]{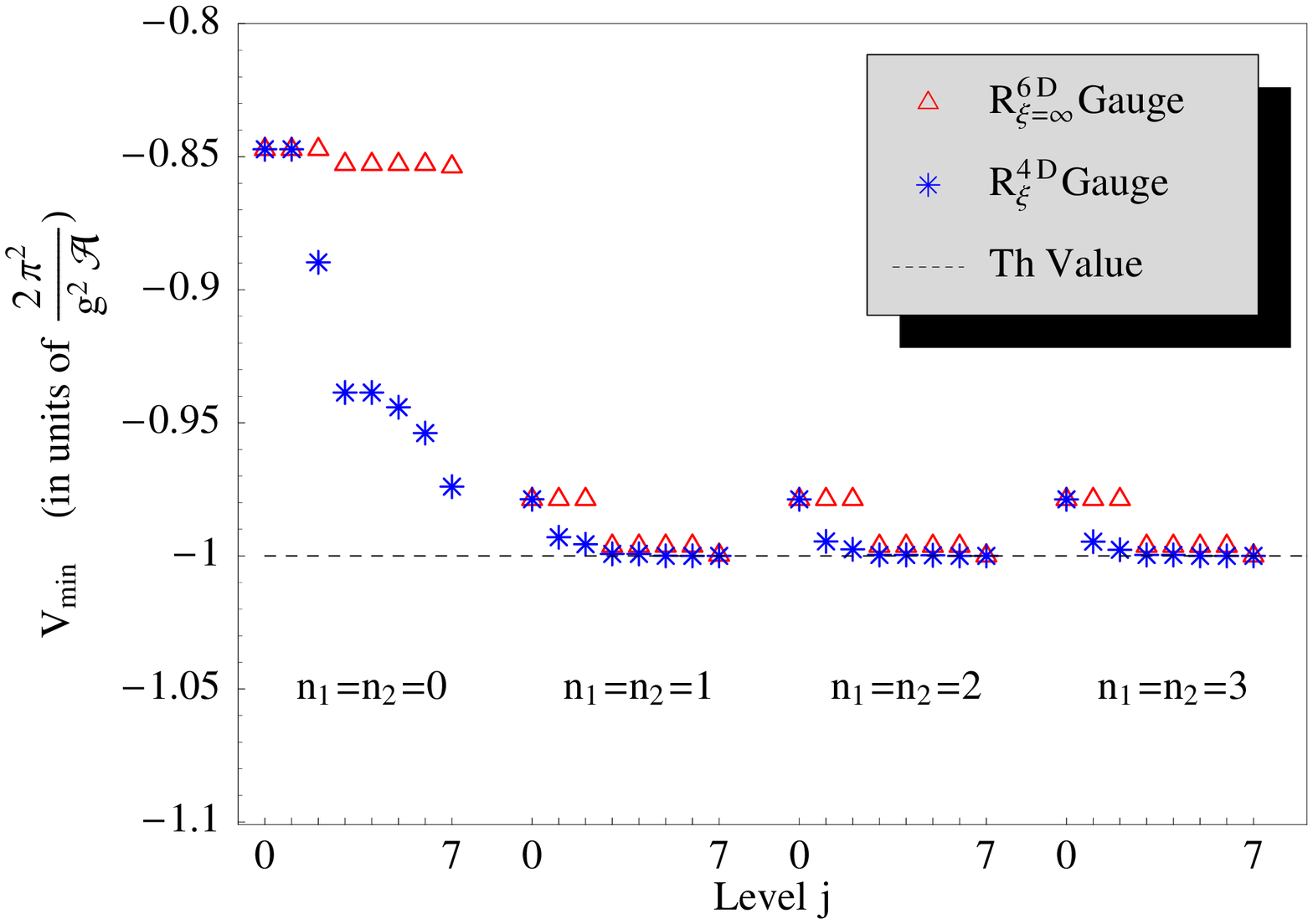}
\caption{\em Values of the minimum of the scalar potential as heavier degrees of freedom are
included. Triangles (stars) represent the numerical results obtained in the $R^{6D}_{\xi=\infty}$
($R^{4D}_{\xi}$) gauge. The horizontal dashed line represents the theoretically predicted value 
for the potential minimum, in the non-trivial 't Hooft flux case.}
\label{vmintotM1}
}

We found that generically all charged and neutral fields in the two towers get {\em vev}s.
Fig.~(\ref{vmintotM1}) shows 
the dynamical approach to the true minimum by the successive addition of heavier 
charged modes (labelled by $j=0,\cdots,7$ in the horizontal axis) and heavier neutral 
modes (labelled with $n_1=n_2=0, \cdots, 3$), 
for both the  $R^{4D}_{\xi}$ and $R^{6D}_{\xi=\infty}$ gauges. For example, 
the point labelled with $n_1=n_2=1$ and $j=3$ represents the numerical calculation where 
{\bf all}  degrees of freedom up to $n_1=n_2=1$ and $j=3$ are included. The graphic shows 
that the value of the minimum of the scalar potential does converge to the theoretically 
predicted value of $-2\pi^2/(g^2 \mathcal{A})$: for $n_1=n_2 \ge 1$ ($\ge 5$ neutral 
complex fields) and $j\ge 3$ ($\ge 4$ charged complex fields) a precision over  
$1\%$ is achieved, in both 
gauges; for $n_1=n_2=3$ and $j=7$, it reaches $10^{-5}$ ($10^{-7}$) for 
the $R^{6D}_{\xi=\infty}$ ($R^{4D}_{\xi}$) gauge.

\FIGURE[t]{
\vspace{-0.5cm}
\includegraphics[width=13cm]{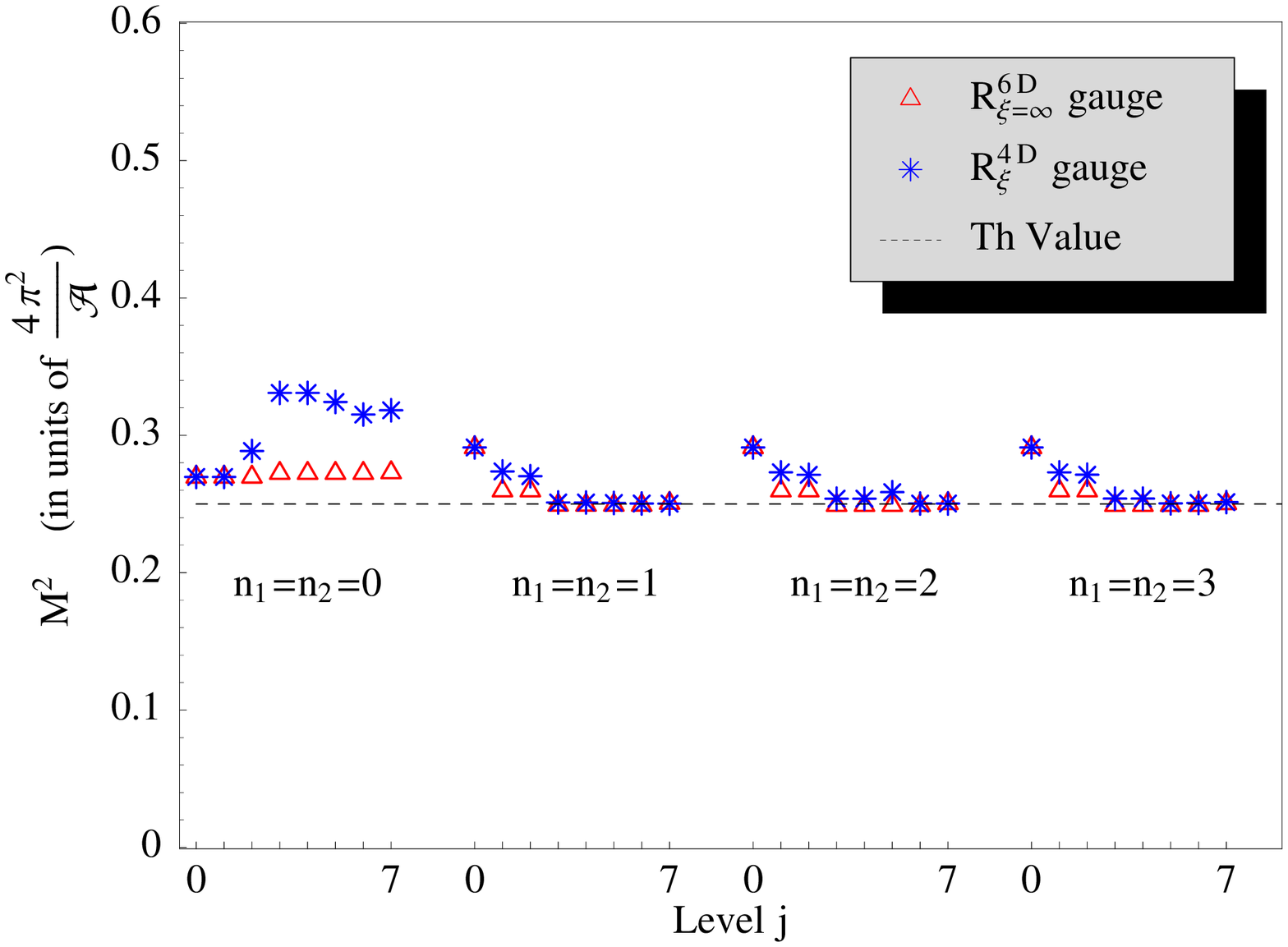}
\caption{\em Lightest gauge mode mass.
Triangles (stars) represent the numerical results obtained in the $R^{6D}_{\xi=\infty}$
($R^{4D}_{\xi}$) gauge. The horizontal dashed line represents the theoretically predicted 
value in the non-trivial 't Hooft flux case.}
\label{MasslessU1M1}
}

As regards the symmetries of the spectrum, the numerical results confirm that the $SU(2)$
symmetry is completely broken. This is well illustrated by Fig.~\ref{MasslessU1M1}, where the 
lightest vector state is shown to be asymptotically massive.  The horizontal 
dashed line represents the mass value of $0.25$ (in units of $4 \pi^2/\mathcal{A}$),  
theoretically predicted in Eq.(\ref{spectrum1_2}). An excellent agreement is observed as well 
between the calculations in the two gauges after the levels up to $n_1=n_2 \ge 1$ and $j\ge 3$ 
are included. We have thus explicitly proved that the $SU(2)$ symmetry is completely broken.

\FIGURE[t]{
\hspace{-0.5cm}
\includegraphics[width=13cm]{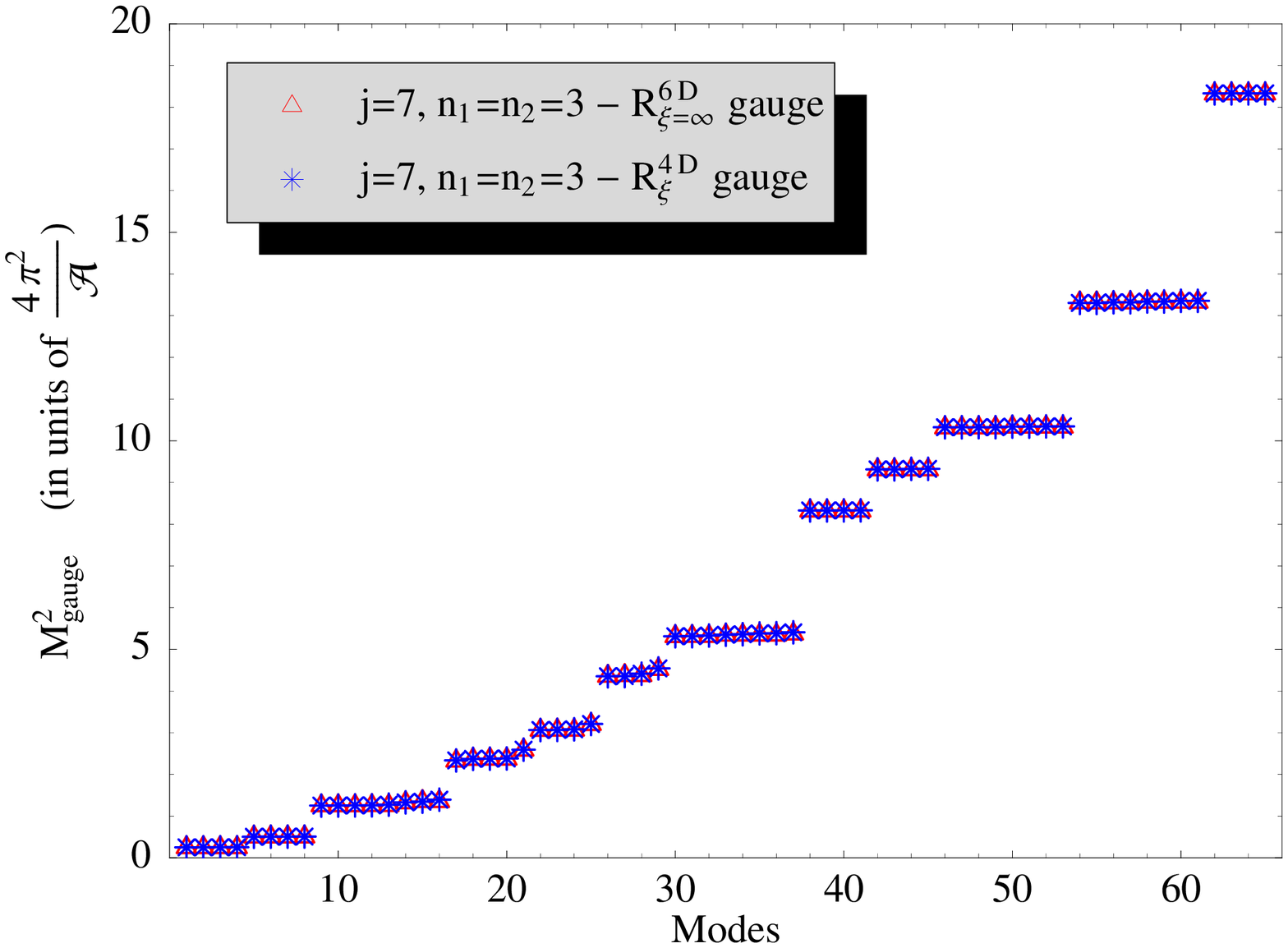}
\caption{\em Gauge invariance of the gauge spectrum for the non-trivial 't Hooft flux case. 
Triangles (stars) represent the numerical results obtained in the $R^{6D}_{\xi=\infty}$ 
($R^{4D}_{\xi}$) gauge respectively, for $n_1=n_2=3$ and $j=7$.}
\label{GauInvM1} }

In Fig.~\ref{GauInvM1} the full spectrum of the $4D$ vector fields is displayed, with all 
fields up to $n_1=n_2=3$ and $j=7$ included in the estimation, in the $R^{4D}_{\xi}$ and 
$R_{\xi=\infty}^{6D}$ gauges. No visible difference can be noticed. This result is a strong 
numerical proof of  the consistency of our effective $4D$ Lagrangian, and its manifest gauge 
invariance when a sufficient number of heavy degrees of freedom are included.

Finally,  Fig.~\ref{ScGauTh73M1} retakes the full spectrum, resulting from the diagonalization 
of the complete system, in the $R^{4D}_{\xi}$ gauge: gauge bosons (stars), physical scalars 
(empty triangles) and unphysical scalars (full triangles), with the latter corresponding to 
the choice $\xi=0$. Superimposed, the Figure shows as well  (black dots joined by a full line) 
the theoretical prediction for  constant discrete Scherk-Schwarz boundary conditions, Eq.(\ref{spectrum1_2}). 
Notice that:
\FIGURE[t]{
\hspace{-0.5cm}
\includegraphics[width=13cm]{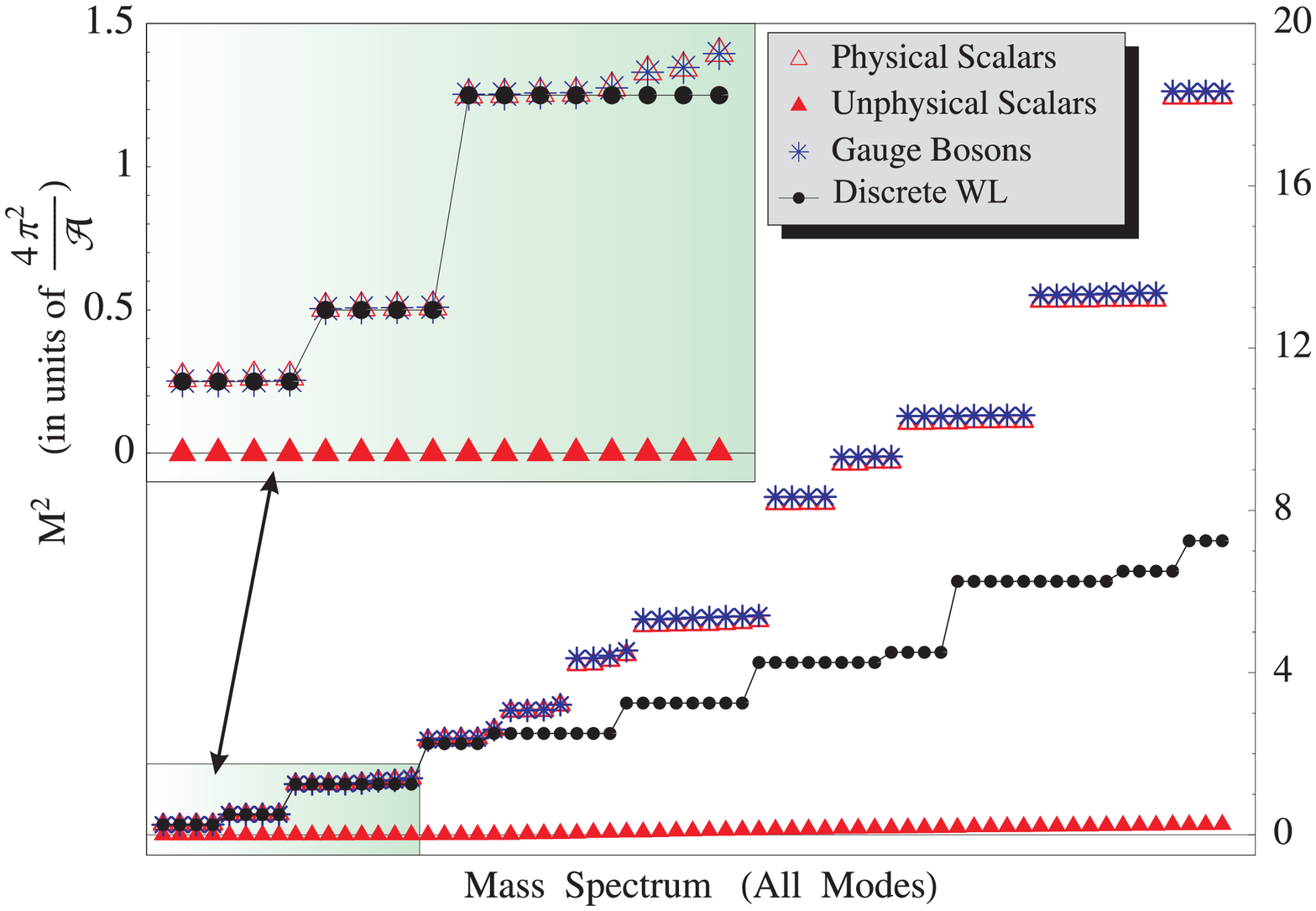}
\caption{\em Full spectrum for the non-trivial 't Hooft flux case, in the $R^{4D}_{\xi=0}$-gauge. 
Gauge bosons (stars), physical scalars (empty triangles) and unphysical scalars (full triangles) 
are shown. The minimization procedure includes all charged and neutral modes up to $n_1=n_2=3$ 
and $j=7$. Black dots joined by a full line represent the theoretically predicted masses derived 
in Section 2.2.}
\label{ScGauTh73M1}
}
\begin{itemize}
\item  Each $4D$ vector boson has a physical scalar partner degenerate in mass,
as expected in the asymptotic limit from Eqs.(\ref{4mgauge}) and (\ref{4mscalar}). 
\item The unphysical scalar spectrum -which constitutes half of the scalar spectrum- 
is identified as those fields which appear to have zero mass, as expected for 
``pseudo-goldstone bosons'' eaten by the vector fields to acquire masses\footnote{As stated, 
this numerical spectrum has been computed for $\xi=0$, but it can also be viewed as 
corresponding to the $\xi$-independent contributions to the goldstone masses for any $\xi$, 
as it follows from Eq.~(\ref{4mscalar}).}. A slight numerical mismatch only appears for the 
masses of the pseudo-goldstone fields of the heavier modes, as the numerical truncation 
of the tower of states starts to be felt.

\item The coincidence between the numerical results -obtained with $y$-dependent boundary 
conditions- and the spectrum predicted for constant discrete Scherk-Schwarz boundary conditions 
(black dots) is very good up to the first 20 modes (i.e. around $M^2\approx 3$ in the 
units chosen for illustration). The agreement of the overall scale, as well as the expected 
four-fold degeneracy of the first two massive levels and the eight-fold degeneracy of 
the next one, are clearly seen. Only the higher levels start to show disagreement with 
the theoretical formulas. This is as it should be, as the present numerical analysis 
was restricted to charged levels up to $j=7$ and neutral ones up to $n_1=n_2=3$. 
Indeed, the next mode non-included in the numerical analysis would be $j=8$, which has 
a squared mass $M^2 \approx 2.7$. In consequence, the numerical results and the theoretical 
prediction start to diverge around this scale. The mode $j=8$ sets the limit of validity of 
the present numerical analysis, while a better agreement can be reached including higher modes.

\end{itemize}

We have also computed the physical spectrum in the $R_\xi^{4D}$ gauge by another procedure: 
the direct substitution of the {\it vev}s obtained from the numerical minimization into 
the {\it total} covariant derivatives in Eqs.~(\ref{4mgauge}) and (\ref{4mscalar}). The 
coincidence with the numerical results shown above is so precise that it would be 
indistinguishable within the drawing precision.

\subsection{Trivial 't Hooft flux: $m=0$, $k=1$}

Consider now the case of trivial 't Hooft flux, in which the generators of the translation 
operators $T_i$ commute. The simplest non-trivial configuration of this type\footnote{That is, 
with lowest degeneracy.} corresponds to $m=0$ and $k=1$. A two-fold degeneracy of the 
charged (Landau) levels is then present, as $d=2$ in Eq.~(\ref{degeneracy}) and $\rho=0,1$.  
In consequence, due to the higher number of states, the numerical treatment is more cumbersome 
than in the previous Subsection.

\FIGURE[t]{
\hspace{-0.5cm}
\includegraphics[width=13cm]{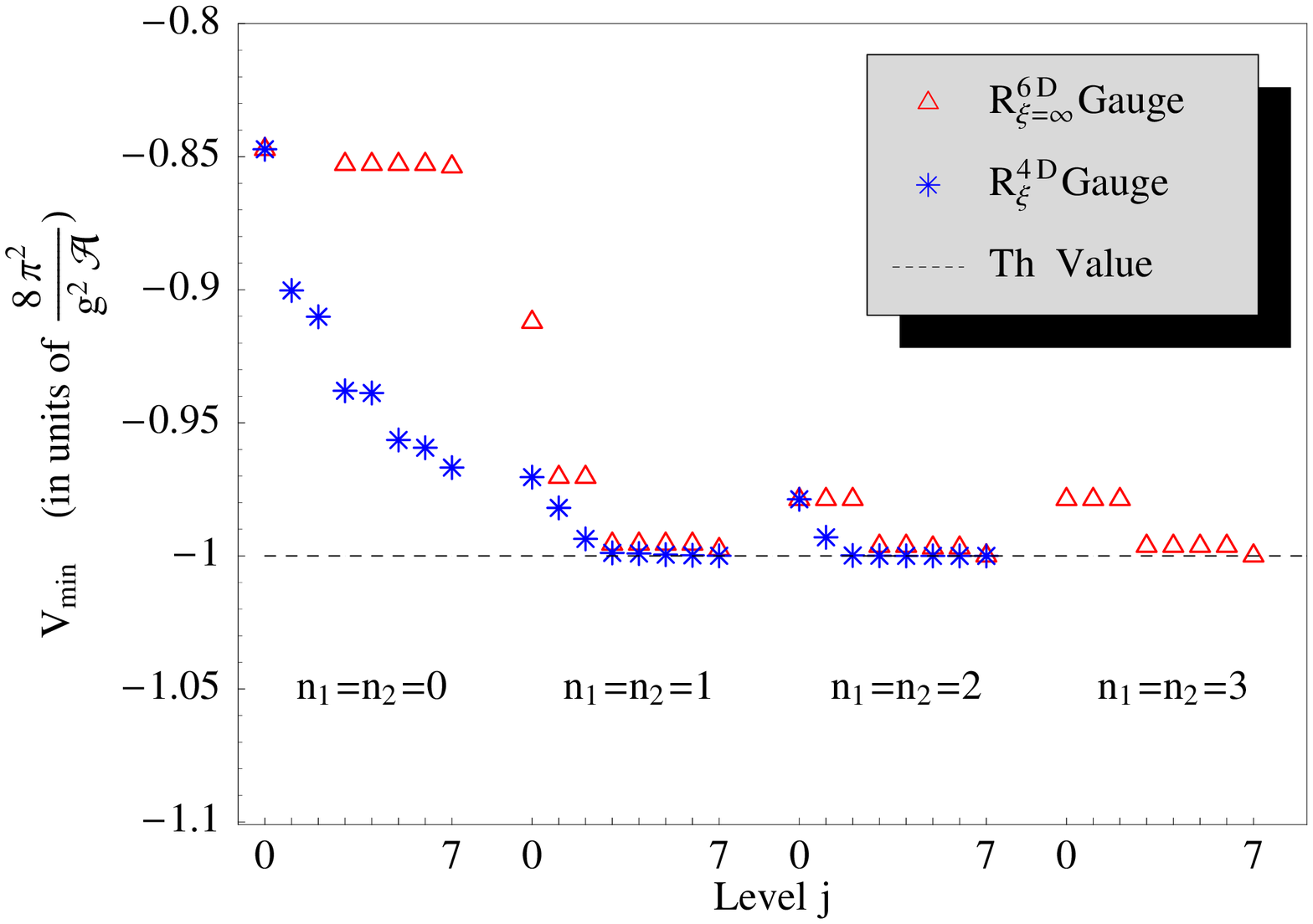}
\caption{\em Values of the minimum of the scalar potential as heavier degrees of freedom are
included. Triangles (stars) represent the numerical results obtained in the $R^{6D}_{\xi=\infty}$
($R^{4D}_{\xi}$) gauge. The horizontal dashed line represents the theoretically predicted value 
for the potential minimum, in the trivial 't Hooft flux case.}
\label{vmintotM0}
}

The dynamical approach to the minimum of the $4D$ potential can be seen in Fig.~\ref{vmintotM0}. 
Again it shows how the asymptotic regime is reached with the successive addition of heavier 
charged and neutral fields. The dashed horizontal line represents the theoretical predicted 
value, $-8\pi^2/g^2 \mathcal{A}$, as expected from Eq.(\ref{vminvalue}): for $n_1=n_2 \ge 1$ 
($\ge 5$ neutral fields) and $j\ge 3$ ($\ge 4$ charged  fields) a precision over  
$1\%$ is achieved, both in the $R^{6D}_{\xi=\infty}$ gauge and in the $R^{4D}_{\xi}$ gauge. 
In the best case that we could numerically evaluate for the $R^{6D}_{\xi=\infty}$ gauge 
($n_1=n_2=3$, $j=7$), a precision of $\mathcal{O}(10^{-5})$ has been obtained.

\FIGURE[t]{
\vspace{-0.5cm}
\includegraphics[width=13cm]{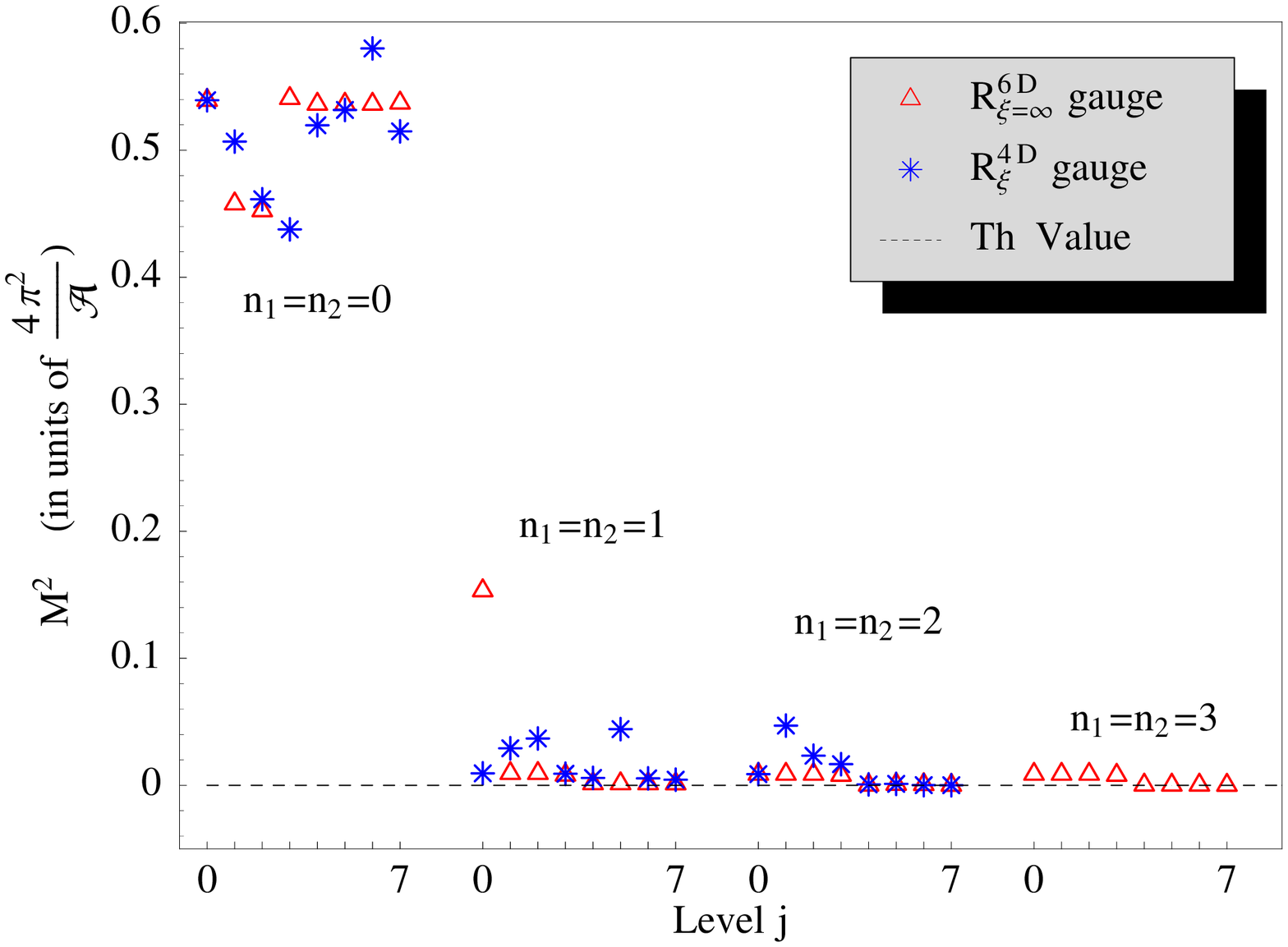}
\caption{\em Lightest gauge mode mass.
Triangles (stars) represent the numerical results obtained in the $R^{6D}_{\xi=\infty}$
($R^{4D}_{\xi}$) gauge. The horizontal dashed line represents the theoretically predicted 
value in the trivial 't Hooft flux case.}
\label{MasslessU1M0}
}

As regards the expected spectra, recall from Subsection 2.1 that all possible solutions 
should correspond to either unbroken $SU(2)$ symmetry or a $SU(2) \to U(1)$ breaking 
patterns, all of them being degenerate in the absence of quantum corrections and fermions.
All numerical results obtained here turn out to correspond to $SU(2) \to U(1)$ breaking 
examples. This is well illustrated by Fig.~\ref{MasslessU1M0} where the mass of one (and only one)  
vector state is seen to vanish asymptotically, in agreement with the lightest value predicted in 
Eq.(\ref{KK}) for $\alpha_i \neq0$. That state is the $4D$ gauge vector boson of the unbroken 
$U(1)$ symmetry. The figure also shows clearly that if only
the first few light levels of the KK and Landau towers would have been considered in the 
analysis, the lightest state would have looked massive, suggesting a fake 
$SU(2) \rightarrow\varnothing$ breaking pattern. Only the inclusion of higher charged and neutral
levels allows to attain the asymptotic regime, unveiling then the remaining $U(1)$ symmetry.
Numerically, the agreement with the theoretical prediction starts to be satisfactory 
for $n_1=n_2 \ge 1$ and $j\ge 3$, analogously to the case with non-trivial 't Hooft flux in 
the previous Subsection.

It is worth pointing out that the $U(1)$ symmetry of the {\it total} stable vacuum selects, 
in general, a different gauge direction, in $SU(2)$ space, than that of the {\it imposed} 
abelian background. In other words, it may be a different $U(1)$ symmetry than that naively 
exhibited by the Lagrangian, when expanded around the {\it imposed} background. The neutral 
and charged towers of fields, as defined 
by the latter, have recombined dynamically, to select the final stable symmetric direction. 
 
\FIGURE[t]{
\vspace{-0.5cm}
\includegraphics[width=13cm]{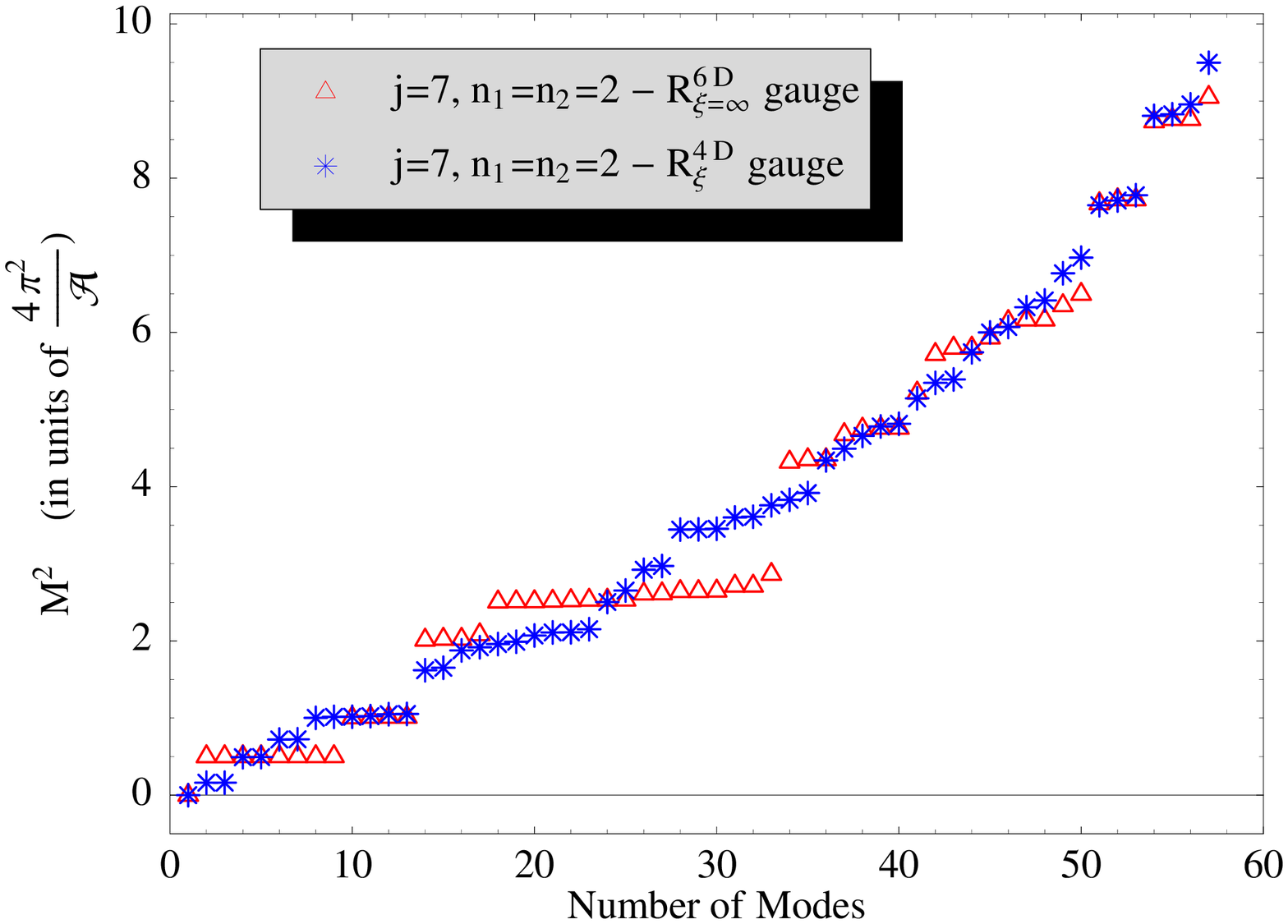}
\caption{\em
Gauge boson spectra for the trivial 't Hooft flux case. Triangles (stars) represent the numerical 
results obtained in the $R^{6D}_{\xi=\infty}$ ($R^{4D}_{\xi}$) gauge respectively, for $n_1=n_2=2$ 
and $j=7$. In this example, the two spectra turn out to correspond to different sets of 
$(\alpha_1, \alpha_2)$ values: $(1/2,1/2)$ (triangles) and $(0.33, 0.22)$ (stars).}
\label{GauInvM0}
}

Fig.~\ref{GauInvM0} shows two gauge spectra obtained numerically including all modes up to $n_1=n_2=2$ 
and $j=7$, for the two gauges $R^{6D}_{\xi=\infty}$ (triangles) and $R^{4D}_{\xi}$ (stars). Notice the 
difference with the analogous figure obtained for the $m=1$ case, Fig.~\ref{GauInvM1}: at first sight, one 
could think that the test of gauge invariance fails in the present case. This is not the case, though: the two spectra turn out to correspond to different values for the set of arbitrary parameters $\alpha_1$, $\alpha_2$, in Eq.~(\ref{spectrum0_2}), which parametrize the possible Scherk-Schwarz spectra.
We determined the values chosen  by the minimization algorithm in these examples, performing a 
two-parameter fit to the first $20$ masses obtained from the numerical procedure. The $\chi^2$ 
value of the fit is extremely significant  for both gauges. It resulted in the values $\alpha_1 = 
\alpha_2 = 1/2$ for the example shown in the $R^{6D}_{\xi=\infty}$ gauge, as can be easily deduced 
from the observed boson multiplicity. Conversely, for the $R^{4D}_{\xi}$ gauge calculation, 
the minimization algorithm selected  $\alpha_1 = 0.334$ and $\alpha_2 = 0.219$, to which it 
corresponds the observed lower multiplicity of degenerate fields. Examples corresponding to 
other values have also been obtained, although not illustrated here. The existence of different 
spectra for the same symmetry breaking pattern is generic of Scherk-Schwarz compactification 
at the classical level. 

\FIGURE[t]{
\vspace{-0.3cm}
\includegraphics[width=13cm]{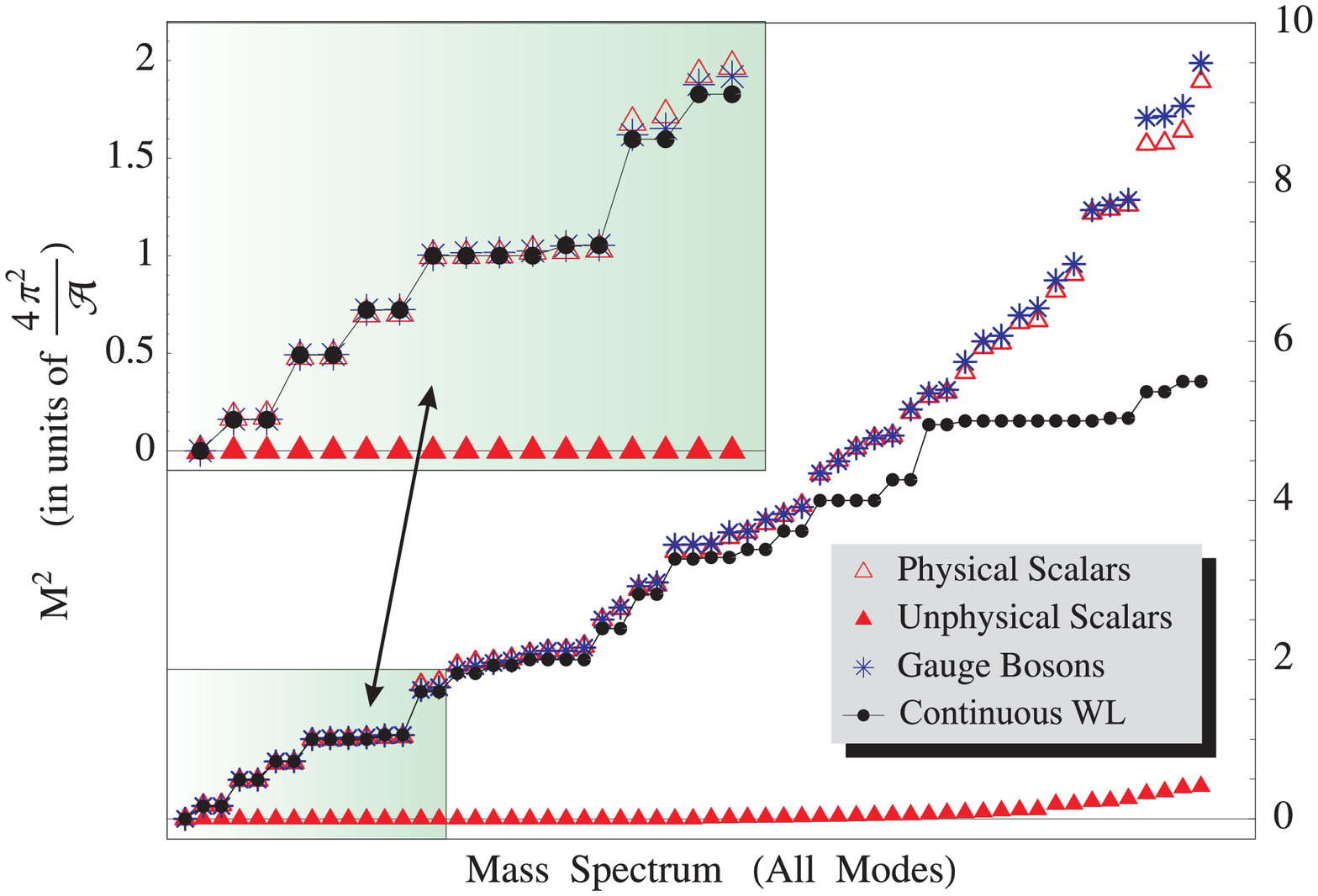}
\caption{\em  Numerical results for the trivial 't Hooft flux case, in the $R^{4D}_{\xi}$-gauge. 
Gauge bosons (stars), physical scalars (empty triangles) and unphysical scalars (stars) are shown. 
The minimization procedure includes all the charged and neutral modes up to $n_1=n_2=2$ and $j=7$. 
Black dots joined by a full line represent the theoretically predicted masses derived in Section 
2.1, for the case $\alpha_1=0.33$, $\alpha_2=0.22$.}
\label{ScGauTh72M0}
}

In Fig.~\ref{ScGauTh72M0} we retake the gauge (stars), physical scalar (empty triangles) and 
unphysical scalar ( full triangles) spectra, in the $R^{4D}_{\xi}$ gauge, for the same $\alpha_i$ 
values than in the previous figure, and with the unphysical scalar masses computed for $\xi=0$. 
Due to the degeneracy of the Landau levels, the numerical analysis could only be performed 
including modes up to $n_1=n_2=2$ and $j=7$. The masses of the unphysical scalar degrees of 
freedom tend, as before, to vanish -as they should- as the asymptotic regime is approached. 
For the heavier modes, a slight numerical mismatch appears between the masses of 
the vector fields and those of their physical scalar partners. A corresponding tiny mass for 
the unphysical scalar partners is also observed. This discrepancy is again consequence of 
the truncation error. Apart form this subtlety, physical scalar and gauge masses are in 
excellent agreement. 

Moreover, the agreement between the numerical spectra and the theoretically 
predicted one - typical of Scherk-Schwarz breaking and represented in Fig.~\ref{ScGauTh72M0} 
with black dots joined by a 
full line - is very good up to the first 40 modes (i.e.around $M^2 \approx 4$ in the units chosen). 
This scale sets the validity limit for the present numerical analysis of our low-energy 
effective $4D$ theory. A better agreement above this scale could be obtained adding higher 
modes. Once again, the mass of the next non-included mode, the $j=8$ mode, is $M^2 \approx 5.4$ 
and coincides with the scale at which the numerical masses and the theoretical predicted ones 
start to diverge.

Finally, we have also computed the physical spectrum in the $R_\xi^{4D}$ gauge by another 
procedure: the direct substitution of the {\it vev}s obtained from the numerical minimization 
into the {\it total} covariant derivatives in Eqs.~(\ref{4mgauge}) and (\ref{4mscalar}). 
The coincidence with the numerical results shown above is so precise that it would be 
indistinguishable within the drawing precision.

\vspace{0.6cm}
In summary, in this Section we have thus explicitly shown, for the $6D$ $SU(2)$ gauge group 
compactified on a $2D$ torus, that 
a stable vacuum of zero energy is reached, out of the initial unstable configuration.
To solve the system with $y$-dependent boundary conditions has been shown to be tantamount 
to solve it with constant boundary conditions. For the case of non-trivial 't Hooft flux, 
the pattern of symmetry breaking obtained is $SU(2) \longrightarrow \varnothing \,$ and 
it corresponds to Scherk-Schwarz symmetry breaking with discrete Wilson lines. 
For trivial 't Hooft flux, the patterns found correspond to $SU(2) \longrightarrow U(1) \,$ 
and are equivalent to Scherk-Schwarz symmetry breaking  with continuous Wilson lines.

\section{Conclusions and outlook}

Boundary conditions depending upon the extra coordinates are equivalent to constant ones, for 
$SU(N)$ on a two-dimensional torus. For trivial 't Hooft flux, they are equivalent 
to constant Scherk-Schwarz boundary conditions, associated to continuous Wilson lines.
For the case of non-trivial 't Hooft flux, the coordinate-dependent boundary conditions 
can be traded instead by constant Scherk-Schwarz boundary conditions, associated to a 
combination of discrete and continuos Wilson lines, resulting always in symmetry breaking. 
One of the novel features of this work is the study of the phenomenological implications 
of this last scenario, studying the pattern of gauge symmetry breaking and the spectrum 
of the four-dimensional vector and scalar excitations.


Chirality cannot be implemented within a $SU(N)$ background and will require to consider in 
the future non-simply connected groups. For them, the equivalence between coordinate-dependent 
and constant boundary conditions does not hold in general. A field-theory treatment of the 
system subject to coordinate dependent boundary conditions is then necessary to solve the 
details of the four-dimensional spectrum. We start this approach in the present work by 
treating also explicitly the case of $SU(2)$ on a torus with background.
  
We have explicitally solved the Nielsen-Olesen instability on the two dimensional torus. 

For the obtention of the four-dimensional effective Lagrangian, all couplings have been 
taken into account, including {\it all} quartic and cubic terms mixing Kaluza-Klein and 
Landau levels. Those terms are shown to be essential in the determination of the stable 
minimum of the potential and its symmetries. The corresponding integrals over the 
extra-dimensional space have been obtained analytically for all modes, for the first time. 
Furthermore, we have defined gauge-fixing Lagrangians, appropiate when both Kaluza-Klein 
and Landau levels are simultaneously present and interacting. We found that the naive 
$R_\xi$ gauge defined in six dimensions is then {\it not} equivalent to the $R_\xi$ gauge 
in four dimensions. The computations have been performed in different possible gauge 
choices and the issue has been clarified in depth. These technical tools will be necessary 
when groups other than $SU(N)$ will be considered.
 
The system is seen to evolve dynamically from the unstable background configuration 
towards a stable and non-trivial background of zero energy. This happens through an infinite 
chain of vacuum expectation values of the four-dimensional scalar fields. The resulting 
spectra do show explicitly the symmetries expected from the theoretical analysis mentioned 
above, for the case of $SU(N)$ with constant boundary conditions.  
 
It turns out that for each four-dimensional gauge boson there exists a scalar partner 
degenerate in mass, both for trivial and non-trivial `t Hooft fluxes. This is one of 
the important phenomenological drawbacks that the approach has to face. The scenario 
has to be enlarged then,  for instance including more than just one scale in the theory.
Indeed, a motivation for the present work was the hypothetical identification of the 
Higgs field as a component of a gauge boson in full space, which would make its mass 
insensitive to ultraviolet contributions, unlike in the Standard Model. To find a 
realistic pattern of electroweak symmetry breaking, which matches the spectra found in 
nature, remains a non-trivial issue.


\section*{Acknowledgments}

We are indebted for very interesting discussions to E. Alvarez, J.
Bellor\'{\i}n, D. Cremades, L.~Ib\'a\~nez and K. Landsteiner.  We are specially
indebted for illuminating discussions and guidance to M.
Garc\'{\i}a-P\'erez, A. Gonz\'alez-Arroyo, E. L\'opez and A. Ramos.
The work of J. Alfaro was partially supported by Secretar\'{\i}a de Estado de
Universidades e Investigaci\'on SAB2003-0238 (Spain) and Fondecyt \# 1060646.
J. Alfaro acknowledges the hospitality of the Departamento de
F\'{\i}sica Te\'orica de la Universidad Aut\'onoma de Madrid (UAM), where
most of this work was done. 
A.~Broncano acknowledges support by the Deutsche Forschungsgemeinschaft within
the Emmy-Noether program. The work of M.B. Gavela, S. Rigolin and M. Salvatori 
was partially supported by CICYT through the project FPA2003-04597 and by CAM 
through the project HEPHACOS, P-ESP-00346. M. Salvatori also acknowledges MECD 
for financial support through FPU fellowship AP2003-1540.
%
%

\appendix
\section{Landau Levels}
\label{Landau_Levels}

In this appendix we derive the wave functions for  the Landau levels on a $2D$ torus \cite{Cremades2}, with charge $q >0$, defined as the solutions of the eigenvalue problem 
\be
\label{eq_mot_+}
a^{\dagger}_{+} a_{+} \,f^{+\,(j)}(y) \,&=& \,j \,f^{+\,(j)}(y) \,, 
\en
where $a_+^\dagger$ and $a_+$ are given in eq.(3.35). They obey the  boundary conditions
\be
f^{+\,(j)}(y \,+\,l_1 ) \,&=& \,e^{i  \pi d \frac{y_2}{l_2}}\,f^{+\,(j)}(y)\,,
\label{l1-per} \\
f^{+\,(j)}(y \,+\,l_2 ) \,&=&\, e^{-i \pi d \frac{y_1}{l_1}}\,f^{+\,(j)}(y) \,,
\label{l2-per}
\en
where $d=q \left( k + \frac{m}{N}\right)$.
It is easy to compute first the zero mode, satisfying
\be
a_+ \,f^{+(j=0)}(y)=0 \,
\label{modozero}
\en
and, subsequently, obtain all the heavier solutions by recursively applying 
the creation operator $a^\dagger$: 
\be
f^{+\,(j+1)}(y) &=& \sqrt{\frac{1}{j+1}} \;a^{\dagger}_+ \;f^{+\,(j)} (y) \,.
\label{jm1}
\en
A possible ansatz for the  wave function $f^{+\,(j=0)}(y)$, compatible with the periodicity condition along  the direction $y_1$ 
in Eq..(\ref{l1-per}), is 
\be
f^{+\,(j=0)}(y) = \sum_{n=-\infty}^{\infty} c_n (y_2) e^{ i \pi d \frac{ y_1 y_2}{l_1 l_2}}  
e^{2 \pi i n \frac{y_1}{l_1}} \;.
\label{prop+}
\en
The periodicity condition along the direction $y_2$,  Eq..(\ref{l2-per}), implies 
that  $d$ {\it must be an \underline{integer}} 
and  the coefficients $c_n(y_2)$ must satisfy the periodicity condition:
\be
c_n (y_2 + l_2) = 
 c_{n+ d} (y_2) \;.
\label{porca}
\en
The coefficients $c_n (y_2)$ are explicitly  obtained after the substitution of  Eq..(\ref{prop+}) into Eq..(\ref{modozero}), giving 
\be
\partial_2 c_n (y_2) = \left( - \frac{2\pi \,d}{l_1 l_2} y_2 - \frac{2 \pi n}{l_1} \right) c_n (y_2) \,,
\label{eqdif+}
\en
with solution  
\be
c_n (y_2) = A_n e^{- \frac{\pi \,d }{l_1 l_2}  y_2^2 - \frac{2 \pi n}{l_1} y_2} \;.
\en
The coefficient $A_n$ is determined by the periodicity condition for the $c_n (y_2)$, Eq..(\ref{porca}), implying
\be
A_{n+ d} = A_n e^{- \pi \frac{l_2}{l_1} (2 n + d )} \,,
\en
whose solution is 
\be
A_n = b_n e^{-  \pi \frac{l_2}{l_1} \frac{n^2}{d}}\,,
\en
where the constants $b_n$  satisfy $b_{n+d} =b_n$. It exists, therefore,  $d$ arbitrary constant coefficients and, consequently, $d$ independent solutions for the zero mode.  We will characterize them 
by the integer number $\rho$, $\rho = 0, ...,  d-1 $, as described in Sect. 3.

All in all, the lightest wave function can be written as 
\be
f^{+\,(j=0)}(y)&=& \sum_{\rho=0}^{d-1} b_\rho \,f^{+\,(j=0,\rho)}(y)\,,
\en
where $b_\rho$ are arbitrary coefficients subject to the normalization condition 
\be
\sum_{\rho=0}^{d-1} |b_\rho|^2 =1\,,
\en
and the functions $f^{+\,(j=0,\rho)}(y)$ are given by  
\be
f^{+\,(j=0,\rho)}(y)=\left(\frac{2 d}{l_1^3 \,l_2}\right)^{\frac{1}{4}} \,\sum_{n=-\infty}^{\infty}  e^{- \frac{\pi d}{l_1 l_2} ( y_2 + n l_2+ \frac{\rho l_2}{ d})^2} 
e^{2 \pi i (d n + \rho) \frac{y_1}{l_1} } e^{ i \frac{\pi d}{l_1 l_2} y_1 y_2} \;.
\label{wavefunctionzero+}
\en
Notice that for  $d>1$  the different independent solutions $f^{+\,(j,\rho)}(y)$ are localized at  different points of the extra dimensions. 

Finally, the expression of  the heavier modes resulting from Eq..(\ref{jm1}) reads:
\be
&&\hspace{-1.5cm} f^{+\,( j,\rho)}(y) = \left(\frac{2 d}{l_1^3 \,l_2}\right)^{\frac{1}{4}}\, \frac{(-i)^j }{\sqrt{2^j\,j!}}\,e^{i  \frac{\pi d}{l_1 l_2}y_1 y_2} \,\nonumber \\
&& \hspace{-0.5cm} \cdot \,\sum_{n=-\infty}^{\infty}\, e^{- \frac{\pi d}{l_1 l_2} (y_2+n l_2+\frac{\rho l_2}{d})^2} \,
e^{2 \pi i \frac{y_1}{l_1} (d n+\rho)} \, H_{j,\rho}\left[ \sqrt{ \frac{ 2\pi d }{l_1 l_2}} \left(y_2+ n l_2 + \frac{\rho l_2}{d} \right) \right]\,, \nonumber \\
\label{paesados}
\en 
with $H_{j,\rho}(y)$  being the Hermite polynomials.

\section{Integrals}
\label{AppIntegrals}

We summarize the integrals of the extra dimensional wave functions, necessary to explicitly obtain  the effective coefficients of the 4D theory.
\begin{itemize}
\item Two-field integrals:
\be
\int_{T^2} \,f^{3\,(n_1,n_2)} \, f^{3\,(m_1,m_2)} d^2 y \,&=&\, \delta_{n_1,-m_1}\,\delta_{n_2,-m_2}\,, \\
\int_{T^2} \,f^{+\,(j_1,\rho_1)} \, f^{-\,(j_2,\rho_2)} d^2 y \,&=&\, \delta_{j_1,j_2}\,\delta_{\rho_1,\rho_2}  \,,
\en
where $f^{3\,(n_1,n_2)}$ and $f^{+\,(j,\rho)}$ are respectively given by eq.~(3.31) and eq.~(3.39).

\item Three-field integrals:

 if $ \frac{\rho_2-\rho_1-n_1}{d} \not\in {\mathbb Z}$, 
\be
I^{(3)}[j_1,\rho_1,j_2,\rho_2,n_1,n_2]= \int_{T^2} \,f^{+\,(j_1,\rho_1)}\,f^{-\,(j_2,\rho_2)} \,f^{3\,(n_1,n_2)}\,d^2 y =0 \,,
\en
else
\be
I^{(3)}[j_1,\rho_1,j_2,\rho_2,n_1,n_2]&=& 
l_1 \,\sqrt{\frac{R}{\mathcal{A}^2}} \,e^{-2 \pi i \frac{\rho_1 n_2 }{d}} \,e^{- \pi i \frac{n_1 n_2}{d}}\, e^{-\frac{\pi}{2 d}(\frac{n_2^2 }{R} + R n_1^2)}\,\frac{\sqrt{j_1! j_2!}}{2^{j_1+j_2}} \,   \\
&& \hspace{-2.cm} \times \sum_{k=0}^{j_2}\,\sum_{k_1=0}^{[\frac{j_1}{2}]}\,\sum_{k_2=0}^{Min[k,j_1-2 k_1]} \, \frac{2^{k_2} (-1)^{k_1} i^{j_1+k-2 k_1-2 k_2}}{k_1! k_2!(j_2-k)!(j_1-2 k_1-k_2)!(k-k_2)!} \, \nonumber  \\
&& \hspace{-2.cm} \times H_{j_1+k-2 k_1-2 k_2}\left[ \sqrt{\frac{\pi}{d}}\left(\frac{n_2}{\sqrt{R}}+i \sqrt{R} n_1 \right)\right]\,
H_{j_2-k} \left[ 2 \sqrt{\frac{\pi R}{ d }} n_1 \right] \,,\nonumber 
\en
where $\mathcal{A}=l_1 l_2$ and $R=l_2/l_1$.

\item Four-field integrals with two charged and two neutral fields:
\be
\hspace{-1.7cm}I^{(4)}_{NC}[j_1,\rho_1,j_2,\rho_2,n_1,n_2,m_1,m_2] &\equiv& \int_{T^2} \,f^{+\,(j_1,\rho_1)}\,f^{-\,(j_2,\rho_2)} \,f^{3\,(n_1\,n_2)}\,f^{3\,(m_1\,m_2)} \,d^2 y  \\
&=&  I^{(3)}[j_1,\rho_1,j_2,\rho_2,n_1+m_1,n_2+m_2] \nonumber \,.
\en

\item Four-field integrals with four charged fields: 

when $\frac{\rho_1+\rho_3-\rho_2-\rho_4}{d} \not\in {\mathbb Z}$, 
\be
\hspace{-2cm}I^{(4)}_{C}[j_1,\rho_1,j_2, \rho_2, j_3, \rho_3, j_4, \rho_4] \equiv  \int_{T^2} \,f^{+\,(j_1,\rho_1)}\,f^{-\,(j_2,\rho_2)} \,f^{+\,(j_3,\rho_3)}\,f^{-\,(j_4,\rho_4)} \,d^2 y =0\,,
\en
else
\be
&& \hspace{-1.9cm} I^{(4)}_{C}[j_1,\rho_1,j_2, \rho_2, j_3, \rho_3, j_4, \rho_4] =  
\frac{\sqrt{d R}}{\mathcal{A}}\,\frac{\sqrt{j_1! j_2! j_3! j_4!}}{2^{j_1+j_2+j_3+j_4}}
\,\sum_{p,k=-\infty}^{\infty} \,e^{-\pi d R \left[(\frac{\rho_1-\rho_2}{d} - k)^2 + (\frac{\rho_1-\rho_4}{d}-p)^2\right]} \nonumber \\
&&\hspace{-0.7cm} \times \sum_{k_1=0}^{j_1}\,\sum_{k_2=0}^{j_2}\,\sum_{k_3=0}^{j_3}\,\sum_{k_4=0}^{j_4}\,\sum_{z_1=0}^{Min[k_1,k_2]}\,
\sum_{z_2=0}^{Min[k_3,k_4]} \,\frac{2^{z_2-z_1+k_1+k_2}(k_1+k_2-2 z_1)! \delta_{k_3+k_4-2z_2}^{k_1+k_2-2 z_1}}{z_1!z_2!(j_1-k_1)!(j_2-k_2)!(j_3-k_3)!(j_4-k_4)!} \nonumber \\
&&\hspace{-0.7cm} \times \frac{H_{j_1-k_1}\left[ -\sqrt{\pi d R}(k+p+\frac{\rho_4+\rho_2-2\rho_1}{d}) \right]\,
H_{j_2-k_2}\left[ -\sqrt{\pi d R}(-k+p+\frac{\rho_4-\rho_2}{d})\right]}{(k_1-z_1)!(k_2-z_1)!(k_3-z_2)!(_4-z_2)!} \nonumber \\
&&\hspace{-0.7cm} \times H_{j_3-k_3}\left[ \sqrt{\pi d R}(k+p+\frac{\rho_4+\rho_2-2\rho_1}{d}) \right]
H_{j_4-k_4}\left[\sqrt{\pi d R}(-k+p+\frac{\rho_4-\rho_2}{d}) \right] \,. \nonumber \\
\en

\end{itemize}

The  integrals above are related by the following completeness relationships, which we have checked  numerically up to a precision better than $10^{-6}$. 
\be
&&I^{(4)}_C[j_1,\rho_1,j_2, \rho_2, j_3, \rho_3, j_4, \rho_4] = \nonumber \\ 
&&= \sum_{n_1,n_2=-\infty}^{\infty} \,\,I^{(3)}[j_1,\rho_1,j_2,\rho_2,n_1,n_2] \,I^{(3)}[j_3,\rho_3,j_4,\rho_4,- n_1, - n_2]\,,  \\
&&\nonumber \\
&&I^{(4)}_{NC}[j_1,\rho_1,j_2, \rho_2,n_1, n_2, m_1, m_2] =  \nonumber \\
&&= \sum_{\rho=0}^{d-1}\, \sum_{j=0}^{\infty}\, I^{(3)}[j_1,\rho_1,j,\rho,n_1,n_2]\,I^{(3)}[j,\rho,j_2,\rho_2,m_1,m_2]\,.
\en

%
%

%
%

\begin{thebibliography}{99}

%
\bibitem{lep}
See for instance [LEP Collaboration], ``A combination of preliminary electroweak 
measurements and constraints on the standard model,'' [arXiv:hep-ex/0312023].
%
\bibitem{technicolour}
L.~Susskind,
Phys.\ Rev.\ D {\bf 20}, 2619 (1979).
%
\bibitem{littlehiggs}
 C.~T.~Hill, S.~Pokorski and J.~Wang,
Phys.\ Rev.\ D {\bf 64}, 105005 (2001);
N.~Arkani-Hamed, A.~G.~Cohen and H.~Georgi,
 Phys.\ Lett.\ B {\bf 513}, 232 (2001);
\bibitem{Manton}
N.~S.~Manton,
Nucl.\ Phys.\ B {\bf 158}, 141 (1979);
D.~B.~Fairlie,
Phys.\ Lett.\ B {\bf 82}, 97 (1979) and
J.\ Phys.\ G {\bf 5}, L55 (1979);
P.~Forgacs, N.~S.~Manton,
Commun.\ Math.\ Phys.\  {\bf 72}, 15 (1980);
N.~V.~Krasnikov,
Phys.~Lett.~B~{\bf 273} (1991) 246.
%
\bibitem{Hosotani}
Y.~Hosotani,
Phys.\ Lett.\ B {\bf 126}, 309 (1983);
Phys.\ Lett.\ B {\bf 129}, 193 (1983).
%
\bibitem{Hatanaka}
  H.~Hatanaka, T.~Inami and C.~S.~Lim,
  Mod.\ Phys.\ Lett.\ A {\bf 13} (1998) 2601.
%
\bibitem{Randjbar}
S.~Randjbar-Daemi, A.~Salam, J.~Strathdee,
Nucl.\ Phys.\ B {\bf 214}, 491 (1983).
%
\bibitem{FineTuning}
  B.~Holdom and J.~Terning, 
  Phys.\ Lett.\ B {\bf 247}, 88 (1990); 
%
  P.~H.~Chankowski, J.~R.~Ellis, M.~Olechowski and S.~Pokorski,
  Nucl.\ Phys.\ B {\bf 544} (1999) 39.
%
  R.~Barbieri and A.~Strumia,
  Phys.\ Lett.\ B {\bf 433} (1998) 63.
%
  J.~A.~Casas, J.~R.~Espinosa and I.~Hidalgo, 
  JHEP {\bf 0401}, 008 (2004); 
%
  J.~A.~Casas, J.~R.~Espinosa and I.~Hidalgo, 
  JHEP {\bf 0503}, 038 (2005).
%
\bibitem{recent}
Comprehensive reviews include:
M.~Quiros, TASI lectures 2002, Boulder {\bf 549-601} (2002), [arXiv:hep-ph/0302189];
C.~Csaki, TASI lectures 2004, [arXiv:hep-ph/0510275]; 
M.~Serone, IFAE 2005, AIP Conf. Proc. {\bf 794}, 139-142 (2005), [arXiv:hep-ph/0508019].
%
\bibitem{orbifold}
L.~J.~Dixon, J.~A.~Harvey, C.~Vafa and E.~Witten,
Nucl.\ Phys.\ B {\bf 261}, 678 (1985) and 
Nucl.\ Phys.\ B {\bf 274}, 285 (1986).
%
\bibitem{Rubakov}
  V.~A.~Rubakov and M.~E.~Shaposhnikov, 
  Phys.\ Lett.\ B {\bf 125}, 136 (1983);
%
  C.~G.~.~Callan and J.~A.~Harvey,
  Nucl.\ Phys.\ B {\bf 250}, 427 (1985).
%
\bibitem{Gross}
D.~J.~Gross, J.~A.~Harvey, E.~J.~Martinec and R.~Rohm, 
Phys.\ Rev.\ Lett.\  {\bf 54}, 502 (1985).
%
\bibitem{tHooft79}
G.~'t Hooft, 
Nucl.\ Phys.\ B {\bf 153}, 141 (1979);
%
\bibitem{Nielsen}
N.~K.~Nielsen and P.~Olesen, 
Nucl.\ Phys.\ B {\bf 144}, 376 (1978);
N.~K.~Nielsen and P.~Olesen, 
Phys.\ Lett.\ B {\bf 79}, 304 (1978);
J.~Ambjorn, N.~K.~Nielsen and P.~Olesen, 
Nucl.\ Phys.\ B {\bf 152}, 75 (1979);
%
\bibitem{Leutwyler}
H.~Leutwyler, 
Nucl.\ Phys.\ B {\bf 179}, 129 (1981).
C.~A.~Flory, 
Phys.\ Rev.\ D {\bf 28}, 1425 (1983)
E.~Elizalde and J.~Soto, 
Annals Phys.\  {\bf 162}, 192 (1985);
G.~Preparata, 
Nuovo Cim.\ A {\bf 96}, 366 (1986);
L.~Maiani, G.~Martinelli, G.~C.~Rossi and M.~Testa, 
Nucl.\ Phys.\ B {\bf 273}, 275 (1986).
%
\bibitem{Cremades}
D.~Cremades, L.~E.~Ibanez and F.~Marchesano,
JHEP {\bf 0405}, 079 (2004).
%
\bibitem{Antoniadis}
I.~Antoniadis, E.~Gava, K.~S.~Narain and T.~R.~Taylor, 
Nucl.\ Phys.\ B {\bf 511}, 611 (1998);
J.~R.~David, 
JHEP {\bf 0209}, 006 (2002).
%
%
\bibitem{Taylor:1997dy}
  For a review see W.~I.~Taylor,
  arXiv:hep-th/9801182.
%
%
\bibitem{Sen:2004nf}
  A.~Sen,
  Int.\ J.\ Mod.\ Phys.\ A {\bf 20}, 5513 (2005).
%
\bibitem{Ambjorn}
J.~Ambjorn and H.~Flyvbjerg, 
Phys.\ Lett.\ B {\bf 97}, 241 (1980);
%
J.~Ambjorn, B.~Felsager and P.~Olesen, 
Nucl.\ Phys.\ B {\bf 175}, 349 (1980).
%
\bibitem{Salvatori}
M.~Salvatori,
 arXiv:hep-ph/0611309 
and
 M.~Salvatori,
 arXiv:hep-ph/0611391.
%
\bibitem{Hebecker}
A.~Hebecker and J.~March-Russell, 
Nucl.\ Phys.\ B {\bf 625} (2002) 128.
%
\bibitem{SS}
J.~Scherk and J.~H.~Schwarz,
Nucl.\ Phys.\ B {\bf 153}, 61 (1979);
Phys.\ Lett.\ B {\bf 82}, 60 (1979).

\bibitem{Ferrara:1988jx}
  S.~Ferrara, C.~Kounnas, M.~Porrati and F.~Zwirner,
  Nucl.\ Phys.\ B {\bf 318}, 75 (1989).

\bibitem{Luscher}
  M.~Luscher,
  Nucl.\ Phys.\ B {\bf 219} (1983) 233.
\bibitem{Witten}
 E.~Witten,
 JHEP {\bf 9802}, 006 (1998). 
%
  M.~Bianchi, G.~Pradisi and A.~Sagnotti,
  Nucl.\ Phys.\ B {\bf 376}, 365 (1992).
M.~Bianchi,
  Nucl.\ Phys.\ B {\bf 528}, 73 (1998)
  [arXiv:hep-th/9711201].
 %
  \bibitem{callan} 
  A.~Abouelsaood, C.~G.~.~Callan, C.~R.~Nappi and S.~A.~Yost,
  Nucl.\ Phys.\ B {\bf 280}, 599 (1987).
%
\bibitem{Guralnik}
  Z.~Guralnik and S.~Ramgoolam,
  Nucl.\ Phys.\ B {\bf 521}, 129 (1998). 
  %
%
\bibitem{vanBaal:1985na}
  P.~van Baal and B.~van Geemen,
  J.\ Math.\ Phys.\  {\bf 27}, 455 (1986);
  B.~van Geemen and P.~van Baal,
  Kon.\ Ned.\ Akad.\ Wetensch.\ Proc.\  B {\bf 89}, 39 (1986).
%
\bibitem{tHooft81}
G.~'t Hooft, 
Commun.\ Math.\ Phys.\  {\bf 81}, 267 (1981).
%
\bibitem{Gonzalez-Arroyo:1982ub}
 A.~Gonzalez-Arroyo and M.~Okawa,
  Phys.\ Lett.\  B {\bf 120} (1983) 174;
 A.~Gonzalez-Arroyo and M.~Okawa,
  Phys.\ Rev.\  D {\bf 27} (1983) 2397.
%
\bibitem{vanBaal:1982ag}
  P.~van Baal,
  Commun.\ Math.\ Phys.\  {\bf 85} (1982) 529.
%
\bibitem{vanBaal:1983eq}
  P.~van Baal,
  Commun.\ Math.\ Phys.\  {\bf 92} (1983) 1.
%
\bibitem{Brihaye:1983yd}
  Y.~Brihaye, G.~Maiella and P.~Rossi,
  Nucl.\ Phys.\  B {\bf 222} (1983) 309.
%
\bibitem{Gonzalez-Arroyo:1997uj}
  A.~Gonzalez-Arroyo,
  arXiv:hep-th/9807108.
%
\bibitem{Hernandez} 
 A. F. Faedo, D. Hernandez, S. Rigolin and M. Salvatori, work in progress.
%
\bibitem{Gava}
  E.~Gava, K.~S.~Narain and M.~H.~Sarmadi, 
  Nucl.\ Phys.\ B {\bf 504}, 214 (1997);
%
\bibitem{Daniel}
  D.~Daniel, A.~Gonzalez-Arroyo, C.~P.~Korthals Altes and B.~Soderberg,
  Phys.\ Lett.\ B {\bf 221} (1989) 136.
%
\bibitem{Giusti:2001ta}
 P.~van Baal, 
 Commun.\ Math.\ Phys.\  {\bf 94} (1984) 397;
 L.~Giusti, A.~Gonzalez-Arroyo, C.~Hoelbling, H.~Neuberger and C.~Rebbi,
 Phys.\ Rev.\ D {\bf 65} (2002) 074506.
%
\bibitem{Cremades2}
  E.~Onofri,
  Int.\ J.\ Theor.\ Phys.\  {\bf 40} (2001) 537
  [arXiv:quant-ph/0007055];
%
 D.~Cremades, private communication;
%
 A.~Gonzalez-Arroyo and A.~Ramos,
  JHEP {\bf 0407}, 008 (2004)
  [arXiv:hep-th/0404022].
%
\end{thebibliography}
\end{document}